\documentclass[sigconf, nonacm]{acmart}
\usepackage{enumitem}
\usepackage{cleveref}

\usepackage{booktabs}   \usepackage{makecell}   \usepackage{pifont}     

\usepackage{listings}
\usepackage{xspace}
\usepackage{tikz}
\usepackage{tcolorbox}

\AtBeginDocument{}

\setcopyright{none} 
\copyrightyear{2026}
\acmYear{2026}
\acmDOI{XXXXXXX.XXXXXXX}

\acmISBN{978-X-XXXX-XXXX-X/XX/XX}

\newcommand{\cmark}{\ding{51}}
\newcommand{\xmark}{\ding{55}}

\newcommand{\parhead}[1]{\vspace{0.2em}\noindent\textbf{#1}}
\newcommand{\ignore}[1]{}

\definecolor{myblue}{rgb}{0, 0, 1}

\newcommand{\papername}{BOOST\xspace}

\Crefname{figure}{Fig.}{Figs.}
\crefname{figure}{Fig.}{Figs.}
\crefformat{section}{#2§#1#3}
\crefformat{subsection}{#2\S#1#3}
\crefformat{subsubsection}{#2\S#1#3}
\newcommand\revhl[1]{\bgroup
  #1\egroup
} 

\newcommand\revarxiv[1]{\bgroup
  \hskip0pt\color{blue!80!blue}#1\egroup
}

\begin{document}

\title[\papername: Concurrent Access to Host Memory and HBM to Accelerate LLM Inference]{\papername: Concurrent Access to Host Memory\\and HBM to Accelerate LLM Inference}
\author{Anish Saxena$^\dagger$}
\affiliation{\institution{Georgia Tech}
  \city{Atlanta}
  \country{USA}
}

\author{Jae Hyung Ju$^\dagger$}
\affiliation{\institution{Georgia Tech}
  \city{Atlanta}
  \country{USA}
}

\author{Hritvik Taneja}
\affiliation{\institution{Georgia Tech}
  \city{Atlanta}
  \country{USA}
}

\author{Po-An Tsai}
\affiliation{\institution{Nvidia Research}
  \city{Westford}
  \country{USA}
}

\author{Aamer Jaleel}
\affiliation{\institution{Nvidia Research}
  \city{Westford}
  \country{USA}
}

\author{Christos Kozyrakis}
\affiliation{\institution{Nvidia Research \& Stanford}
  \city{Santa Clara}
  \country{USA}
}

\author{Moinuddin Qureshi}
\affiliation{\institution{Georgia Tech}
  \city{Atlanta}
  \country{USA}
}

\begin{abstract}

GPU memory bandwidth and capacity limit throughput in large language model (LLM) inference.
The GPU memory system consists of a primary tier of high-bandwidth memory (HBM) and a secondary tier of host memory connected via CPU-to-GPU interconnect.
Current serving systems treat the tiers hierarchically: they serve exclusively from HBM when data fits, and otherwise prefetch data from host memory to HBM before use.
In both cases, the host memory bandwidth is never well utilized.
Prefetching expands capacity by utilizing host memory, but consumes HBM bandwidth for writes, reducing the bandwidth available for demand loads.
We observe that fully utilizing both host and HBM bandwidth requires each wave of GPU threadblocks to access both tiers \textit{concurrently} and in \textit{proportion} to their bandwidth ratio.
Existing bandwidth-proportional placement strategies fail to provide concurrency because they are not aware of GPU waves, and the large 2MB GPU page size.
  
  This paper presents {\em{\papername}}, the first runtime system that provides concurrent and proportional access to both GPU memory tiers, extracting the combined bandwidth of host memory and HBM for LLM inference \textit{without kernel changes}.
  The key insight in \papername is to use kernel access patterns to make page allocation and runtime data management \textit{wave-aware}.
    For static model weights, \papername applies modulo-based page placement that eliminates access-ratio variance; for dynamically provisioned attention key-value (KV) pairs, it makes the free KV page pool wave-aware.
  We integrate \papername into vLLM and evaluate on a Grace Hopper system. At iso-batch size, \papername improves Time-per-Output-Token (TPOT) by 4.3\% over HBM-only serving, whereas prefetching degrades TPOT by 6\%. In high-throughput serving, \papername improves throughput by 31\% on average, outperforming prefetching by 15\%.

\end{abstract}

\maketitle
\section{Introduction}

The memory system of the GPU determines the processing speed of modern Large Language Models (LLMs)~\cite{acharya2025agentic, gpt4_moe, deepseekv3, zhong2024distserve}.
The decode phase of LLM inference is autoregressive, where hundreds of gigabytes of model weights and attention key-value (KV) pairs are fetched from GPU memory at every iteration, making LLM decoding limited by memory bandwidth, and often capacity.
\let\thefootnote\relax\footnote{$^\dagger$ Equal contribution. Anish and Jae can be 
reached at \{asaxena317,jhju\}@gatech.edu}

The GPU memory system consists of two tiers. The first tier is {\em High-Bandwidth Memory (HBM)}, the primary tier for serving memory requests.
The second tier is {\em host} memory located at the CPU and connected to the GPU via a coherent {\em CPU-to-GPU (C2G)} interconnect. Host memory bandwidth is lower than HBM bandwidth, but is often a significant fraction of it.
For example,
Nvidia's Grace Hopper~\cite{fusco2024understanding, gh200_whitepaper} has host memory with approximately 9\% to 11\% of the HBM bandwidth depending on SKU. Therefore, the overall memory bandwidth of a system that utilizes both HBM and host memory is approximately 1.1x that of an HBM-only system.

Ideally, the system should exploit the combined bandwidth of both tiers, improving the {\em Time-Per-Output-Token (TPOT)} by up to 1.1x. The host tier also adds capacity to batch more requests and raise system throughput.
For instance, if a system uses 20\% of the HBM capacity to store KV cache, then 10\% additional capacity from the host tier increases the KV space by 50\%, allowing up to 1.5x more concurrent requests and up to 1.5x higher throughput.

\begin{figure*}
    \centering
     \includegraphics[width=0.99\textwidth]{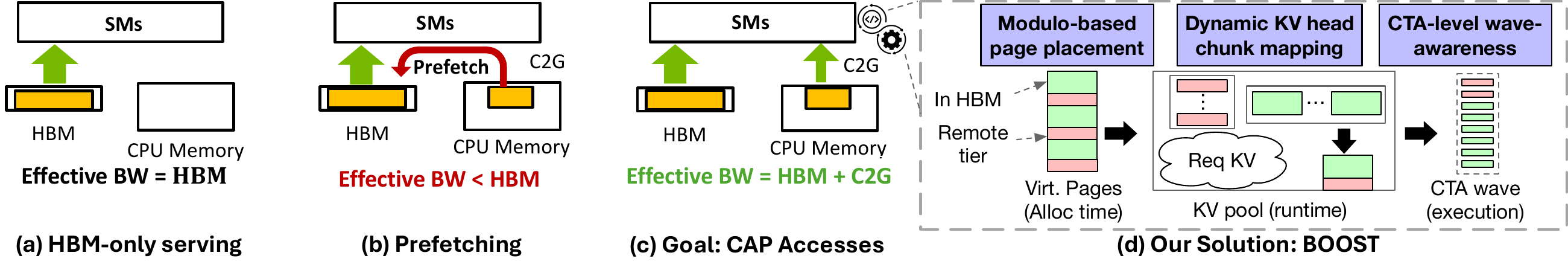}
      \vspace{-3mm}
     \caption{(a) LLM serving systems use HBM exclusively, leaving host CPU memory idle.
     (b) To improve serving capacity, current systems migrate (prefetch) data, degrading HBM demand bandwidth.
     (c) Our goal is to perform concurrent and proportional (CAP) accesses at each wave to fully utilize system bandwidth.
     (d) \papername facilitates CAP accesses by making page placement and runtime data management aware of CTA-level kernel data access patterns.}
    \label{fig:intro}
    \vspace{-2mm}
\end{figure*}

Existing systems do not fully utilize the available memory bandwidth and capacity.
\cref{fig:intro}-(a) shows that when HBM capacity is sufficient to hold the model weights and KV cache, data is served exclusively from HBM, leaving host memory idle.
When the working set exceeds HBM capacity, as \cref{fig:intro}-(b) shows, current systems~\cite{cao2025moe-lighning, yuzuguler2025preserve} place a portion of the data in the host memory and migrate (prefetch) the data from host memory to HBM before use.
However, migrating data from the host to HBM consumes HBM write bandwidth, reducing the available HBM read bandwidth that serve GPU demand loads.
Even an ideal prefetcher that fully overlaps migration delivers less read bandwidth than HBM alone, because every prefetching incurs HBM writes.
The fundamental problem is that current systems treat the two tiers as a hierarchy rather than peers, and data is directly served only from HBM.

To fully extract system-wide memory bandwidth, data accesses must satisfy two properties: {\em{concurrency}} and {\em proportionality}. The first property dictates that accesses must concurrently go to both tiers, even over short timescales.
The second property dictates that accesses be split proportionally to the bandwidth ratio of each tier. For example, if host memory has 10\% of HBM bandwidth, it should receive 10x fewer accesses than HBM.
We note that violating the second property can cause slowdown compared to HBM-only serving. For example, if more than a proportional share of accesses go to host memory, the system would become bottlenecked by host bandwidth while HBM bandwidth remains underutilized.

Prior works on heterogeneous GPU memory~\cite{agarwal2015unlocking, chou2017batman, agarwal2015page}
achieve proportionality by randomly assigning pages across tiers in proportion to their bandwidth ratio. Developed a decade ago for 4KB pages, these schemes obtained concurrency implicitly: a wave touched thousands of pages, so a proportional share reached the host tier within every wave. Modern GPUs use 2MB pages to limit TLB pressure, so a wave touches only tens of pages, and random placement that is proportional on average deviates widely within a wave, which underutilizes the host tier or, worse, stalls on it and idles the HBM.
Our evaluations show that such a random page allocation policy does not benefit LLMs.
Deployed serving systems, meanwhile, use the host tier only through prefetching; serving inference directly from data placed across both tiers remains unexplored.

The goal of our work is to enable {\em Concurrent and Proportional (CAP)} memory accesses to both tiers of the GPU memory system, as shown in \cref{fig:intro}-(c).
However, LLM serving systems are complex, with multiple abstraction layers between physical memory and the runtime data accessed by kernels, which render page-to-data mapping opaque to the kernel and complicate CAP accesses. Moreover, making low-level modifications to highly optimized kernels is both intrusive and impractical, as some kernels are closed source~\cite{cublas}.
Ideally, we want to enable CAP without requiring kernel changes.

In this paper, we present \textit{\papername}, the first runtime system that enables CAP memory access, extracting the additive bandwidth of both host and HBM tier for LLM inference, \textit{without} any kernel modifications.
\papername distributes data to both memory tiers, thus also increasing the usable memory capacity to maximize system throughput.
The core idea behind \papername is to leverage access patterns in key LLM kernels to guide page placement and runtime data management, 
facilitating CAP transparently to the kernel (\cref{fig:intro}-(d)). \papername overcomes three key challenges.

\noindent{\bf{First, page placement.}}  Each wave (set of concurrently executing threadblocks) accesses tens of megabytes of data, spanning only tens of 2MB pages, so random allocation varies the access ratio across waves.
We want \textit{deterministic} access ratio to the two tiers even with a few pages.
We find that kernels that use static data, such as model weights, access virtually contiguous memory during its waves.
Therefore, we propose {\em{Modulo-Based Page Placement (MPP)}}: for an HBM-to-host bandwidth ratio of $K$, MPP allocates the first $K$ pages to HBM and the $(K+1)^{th}$ page to host memory.
MPP enables CAP for static data even with a few 2MB pages.
 
\noindent{\bf{Second, granularity.}}  Kernel launch parameters (warps per threadblock and total threadblocks) vary significantly.
We could guide MPP using data access patterns at the thread, warp, or threadblock level (also known as cooperative thread arrays, or CTAs). Our characterization on GH200 GPU reveals that CTA size shows little variance (4 or 8 warps) but warp specialization (different warps in a CTA access different tiers) is too coarse: 4-warp CTAs only support bandwidth ratios in multiples of 0.25.
Thread specialization is ineffective because memory divergence prevents request coalescing.
Our second insight is to use {\em{CTA-level wave-awareness}} to guide data management, enabling CAP for typical LLM kernels.

\noindent{\bf{Third, handling dynamic data.}} CTAs in attention kernels access dynamically allocated KV cache.
By default, the serving system adds and removes KV pages~\cite{pagedattention_vllm} from the pool of free KV pages at runtime without being wave-aware.
Over time, the dynamic allocation and freeing of KV pages make the mapping from virtual address to CTAs near-random, which no static placement can correct.
Moreover, recent attention kernels~\cite{ ye2025flashinfer, shah2024flashattention} dynamically partition the KV cache of each request into chunks when mapping KV to CTAs.
Because the chunk boundaries change over time, it is challenging to find a static data placement across tiers that achive wave-awareness at CTA-level.
Our third insight is that, despite the dynamism, chunks often align with KV head and request boundaries due to the inherit parallelism. By mapping certain KV head of selected requests to the host tier, we can approximately maintain the desired memory ratio for attention kernels.

We integrate \papername into vLLM and evaluate three LLMs on a Grace Hopper system with a host to HBM bandwidth ratio of 10\%.
In iso-batch low-latency serving, \papername improves TPOT by 4.3\% over HBM-only serving, while prefetching degrades TPOT by 6\%. In high-throughput serving, \papername improves throughput by 31\% on average: the added host capacity provides most of this gain, and concurrent access contributes a further 4\% that no capacity-only mechanism provides. \papername outperforms prefetching by 15\% owing to its higher demand load bandwidth.

\smallskip
Overall, our paper makes the following contributions:
 
\begin{itemize}
 
    \item We make the key observation that performance of LLM serving can be improved by exploiting the bandwidth of both HBM and host memory. To do so, we need \textit{Concurrent and Proportional} (CAP) accesses to both tiers. We show that prior placement schemes provide proportionality but lose concurrency when GPUs moved from 4KB to 2MB pages.
 
    \item We present \papername, the first solution that enables CAP accesses for LLM serving through wave-aware modulo page placement for static model weights and wave-aware runtime management for dynamic data, \textit{without kernel changes.}
 
    \item We show that \papername provides a 4.3\% TPOT improvement at iso-batch and a 31\% average throughput speedup over HBM-only serving (with 4\% owing to concurrent accesses to both tiers), outperforming prefetching by 15\%.
\end{itemize}

\clearpage
\section{Background and Motivation}

\subsection{Large Language Model Inference}
 
Large language models (LLMs) dominate modern GPU workloads. The decode phase
generates tokens autoregressively and reloads the model weights every
iteration, so it dominates end-to-end latency, increasingly so for reasoning
models~\cite{muennighoff2025s1}. Decoding is memory bandwidth-bound: attention
and MLP kernels for dense LLMs, and expert kernels for Mixture-of-Experts
(MoE) models, repeatedly fetch the KV cache and model weights. DeepSeek
R1~\cite{guo2025deepseek}, for instance, holds 670GB of weights, and its KV
cache adds 1GB per user at 32K context. Serving systems~\cite{vllm, trtllm,
tgi, zheng2024sglang, gim2025pie} layer kernel and scheduling
optimizations~\cite{ye2025flashinfer, shah2024flashattention,
agrawal2024taming, pagedattention_vllm} between a kernel and its demand loads,
so the memory system observes complex access patterns. \revhl{We target the
bandwidth-bound decode phase under prefill-decode (PD)
disaggregation~\cite{zhong2024distserve}}, in both low-latency (minimizing TPOT) and
high-throughput (maximizing TPS) serving; the latter is also capacity-bound.

\subsection{Kernel Execution on GPUs}
 
GPUs execute kernels using a hierarchical scheduling model that maps computation to the streaming multiprocessors (SMs). The kernel is divided into many threadblocks (also known as cooperative thread arrays, or CTAs), depending on problem size. Each CTA comprises several warps of 32 threads. The GPU scheduler assigns multiple CTAs to SMs depending on available resources (register files, shared memory, and warps), and once assigned, a CTA occupies SM resources until it completes execution.
Within each active threadblock, the SM can dynamically schedule warps in and out of execution based on their readiness.
LLM kernels can launch thousands of threadblocks, and typically, hundreds of threadblocks are active at each cycle. We refer to the set of active threadblocks across all SMs in the GPU as a \textit{wave}.

\subsection{GPUs with Heterogeneous Memory Systems}
 
\begin{figure}[htb!]
    \centering
    \includegraphics[width=0.7\linewidth]{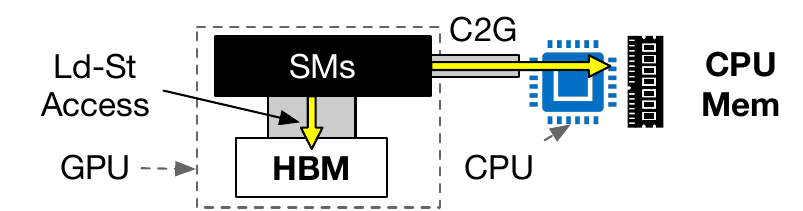}
    \caption{The GPU memory system connects HBM and host CPU memory (attached via C2G links) to the GPU in parallel.}
    \label{fig:hmem_system}
    \vspace{-3mm}
\end{figure}

GPU memory systems are heterogeneous, comprising two memory tiers: the primary local memory (typically HBM) and a secondary host memory tier accessible through CPU-to-GPU (C2G) interconnect~\cite{CAQA25}. As shown in \cref{fig:hmem_system}, GPUs have access to CPU memory through cache-coherent C2G links with load-store semantics. However, the secondary tier provides only a fraction of HBM's bandwidth.
We define $\alpha$ as the ratio of C2G read bandwidth to HBM bandwidth. In Nvidia's Grace Hopper (GH200) system~\cite{fusco2024understanding, gh200_whitepaper}, $\alpha$ ranges from 9\% to 11\% as C2G read bandwidth is 450GB/s and HBM bandwidth ranges between 4TB/s to 4.9TB/s depending on the SKU.
Unlike traditional PCIe-connected systems where one CPU drives multiple GPUs, we focus on tightly-coupled CPU-GPU systems like GH200 with a measured bandwidth ratio of 10\% (\cref{sec:insights} to \cref{sec:eval}). We further analyze future systems with different $\alpha$ (\cref{sec:bw_ratio_sens} to \cref{sec:future}).

\subsection{Requirements for Additive Bandwidth}
\label{sec:requirements}

To fully utilize the HBM+C2G bandwidth ($1+\alpha$), accesses from SMs must satisfy two
properties.
\textit{Proportionality} requires that
$\alpha$ ratio of accesses reach host relative to HBM~\cite{agarwal2015unlocking, chou2017batman}.
\textit{Concurrency}, a property we identify in this work, requires that the accesses within \textit{every} wave reach both tiers in proportion.
We show that violating either property impedes performance.
Existing schemes which use host tier are either \textit{migration-based} that move data to HBM before use, or \textit{placement-based} that serve data from the tier where it resides.

\subsection{Page Migration Schemes}

Migration-based schemes stage data from the host tier in HBM before the SMs consume it, in two forms: prefetching in serving systems, and fault-driven migration in CUDA unified memory.

\smallskip
\noindent\textbf{Prefetching.}
When the model and KV cache exceed HBM capacity, serving systems offload data to host memory and
prefetch it to HBM before use~\cite{vllm_prefetching, cao2025moe-lighning, yuzuguler2025preserve,
hu2025tightllm, pan2025instattention}. Prefetching enables larger workloads to fit in the system but
degrades demand bandwidth. Each byte read from host memory incurs a write to HBM, consuming bandwidth
on the HBM interface.
The writes to HBM are recurrent, as the offloaded data exceeds the HBM staging buffer capacity, so staged bytes are evicted after use and fetched again at the next decode iteration.
The writes reduce effective demand bandwidth from the ideal $(1+\alpha)$ to $(1-\alpha)$; even an ideal prefetcher that fully overlaps migration with computation cannot exceed this bound. In contrast, concurrently loading from both tiers attains $(1+\alpha)$ demand bandwidth, a $(1+\alpha)/(1-\alpha)$ advantage over any prefetcher. At $\alpha=10\%$, this benefit is 22\%, and grows with increasing $\alpha$ (\cref{sec:future}).

\begin{figure}[!htb]
    \centering
    \includegraphics[width=\linewidth]
    {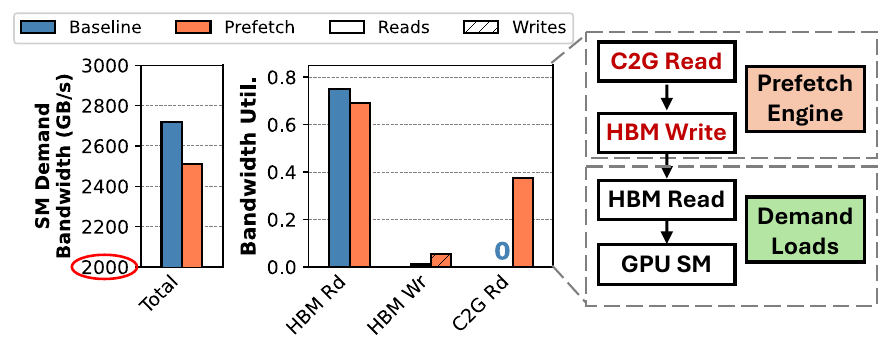}
    \vspace{-5mm}
    \caption{Left: Demand bandwidth to the SM in HBM-only baseline and prefetching for Llama-3.3 70B on GH200, both at batch size of 10 (details in~\cref{sec:eval}). Middle: Breakdown of system bandwidth utilization. Right: Lifecycle of offloaded data. The prefetch engine migrates data from the C2G tier to HBM before use, reducing demand bandwidth available to the SMs.}
    \vspace{-3mm}
    \label{fig:motivate_prefetch}
\end{figure}

\smallskip
\noindent\textbf{Impact of Prefetching on Demand Bandwidth.} \cref{fig:motivate_prefetch} shows the system bandwidth utilization for Llama-3.3 70B running on a Grace Hopper system using vLLM~\cite{vllm_prefetching} (evaluation details in~\cref{sec:eval}). The serving system offloads one out of every $K$ layers and prefetches it asynchronously before use, where $1/K$ approximates the bandwidth ratio required to hide the migration latency from the host tier to HBM. Prefetching to HBM leads to two problems: (i) the migration degrades the HBM read bandwidth available for demand accesses (by 211GB/s in \cref{fig:motivate_prefetch}-left), and (ii) staging requires a buffer in HBM, marginally reducing usable capacity.
To maximize performance, a system must launch demand accesses (not prefetches) to both memory tiers, rendering data migrations unnecessary.

\smallskip
\noindent\textbf{CUDA Unified Memory.}
CUDA performs automatic migration via \texttt{cudaMallocManaged} that moves pages on faults toward the accessing processor to optimize locality. 
Unfortunately, once the working set (model+KV state) exceeds HBM, such a demand-based migration demotes the HBM into a page cache which gets thrashed on every iteration, because each page access brings in the faulted page and evicts another page. 
This makes decoding effectively a sequential on-demand fetch of non-HBM-resident pages, degrading performance significantly---we measure a 5$\times$ throughput degradation (evaluations in \cref{sec:eval_baselines}).  The migrations also consume HBM write bandwidth similar to prefetching.
Both migration schemes serve every byte from HBM and do not issue demand loads to the host tier, so they provide neither proportionality nor concurrency.

\subsection{Page Placement Schemes}
\label{sec:limitations}
\label{sec:os_cuda}

Placement schemes instead treat the two tiers as peers, and data resides where it is placed while the SMs issue demand loads to both tiers, so no migration occurs. Note that serving systems use the host tier only through prefetching (\cref{fig:motivate_prefetch}). No prior work applies placement-based serving to LLM inference. We implement both schemes of this family and present its first characterization.

\smallskip
\noindent\textbf{Bandwidth-Aware Page Placement.}
Prior works on data management for heterogeneous memory~\cite{agarwal2015unlocking, chou2017batman, agarwal2015page} aim to achieve proportionality through driver-level bandwidth-aware page placement, which distributes pages so a randomly selected $\alpha$ fraction resides in the host tier (BAPP-R).
BAPP-R was developed a decade ago for GPUs that used small pages (e.g., 4KB) and obtained concurrency implicitly, because a wave typically accesses tens of megabytes and therefore touches thousands of 4KB pages.
However, modern GPUs use 2MB pages to limit TLB pressure, and BAPP-R fails to achieve concurrency even with bandwidth-proportional placement.
No implementation of BAPP-R exists for modern GPUs or systems, so we implement it ourselves.
To characterize placement, we use a controlled microbenchmark: an add kernel that streams 1GB through thousands of CTAs across several waves, emulating the access pattern of LLM attention, expert, and MLP kernels (\cref{sec:tb_specialize}) with trivial compute, so any bandwidth loss is attributable to placement alone.

\begin{figure}[!htb]
    \centering
    \includegraphics[width=0.7\linewidth]{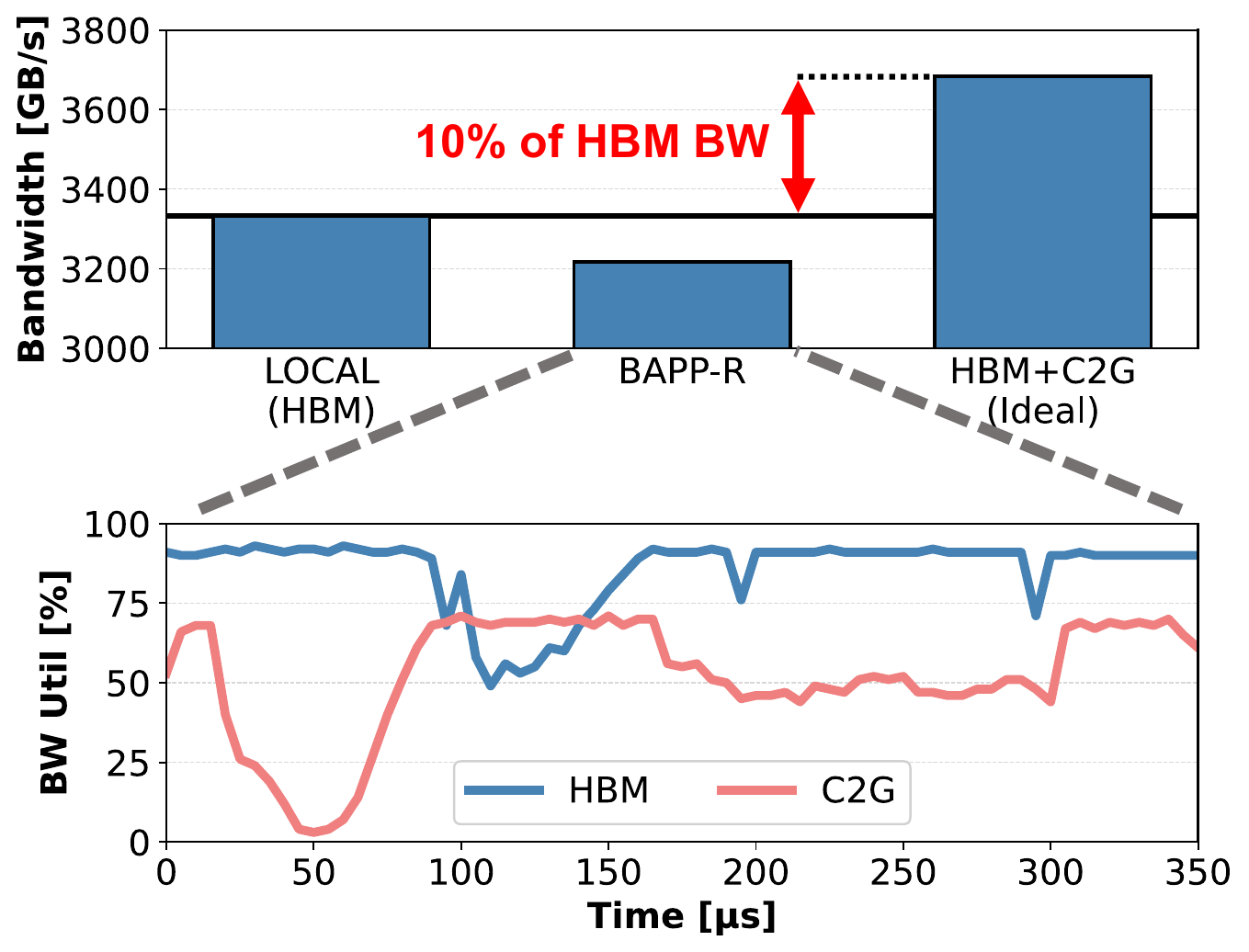}
    \vspace{-3mm}
    \caption{Top: Bandwidth usage of random page placement (BAPP-R) over HBM-only accesses. BAPP-R incurs a 3.5\% lower usage while the ideal bandwidth improvement is 10.4\%. Bottom: Bandwidth utilization of HBM and C2G links over time for BAPP-R. The temporal variance in bandwidth utilization degrades system bandwidth utilization.}
    \label{fig:motivation_combined}
    \vspace{-3mm}
\end{figure}

\smallskip
\noindent\textbf{Impact of BAPP-R on Demand Bandwidth.}
\cref{fig:motivation_combined} shows that BAPP-R
degrades demand bandwidth to SMs by 115GB/s (3.5\%) below the HBM-only baseline, against an ideal
improvement of 10.4\%.
We find that BAPP-R guarantees proportional access only on average over the full kernel,
not within each wave, which creates variable bandwidth ratios \textit{across waves}. 
To illustrate, consider a biased
coin with $P(\text{heads}) = 10\%$: across hundreds of flips roughly 10\% are heads, but windows of few
tens of flips deviate significantly. BAPP-R behaves the same. Across the full kernel (hundreds of 2MB
pages) the placement ratio is $10\% \pm 1.5\%$ over 100 kernel instances); within a wave it varies far more,
$10\% \pm 3.6\%$ (over 2,000 waves).
The variance is asymmetrically harmful. Waves that over-access HBM
underutilize the C2G link, and, more critically, waves that over-access the host tier stall on its
lower bandwidth and underutilize HBM, degrading overall system bandwidth utilization.
Reverting to 64KB pages does not work because the resulting TLB misses 
degrade HBM bandwidth below 1000GB/s, less than one-third of its maximum bandwidth (not shown in figure).
Deterministic page interleaving could fix this randomness-induced variance.

CUDA does not expose bandwidth-aware allocation: \texttt{cudaMalloc} allocates only on HBM, the Virtual Memory Management (VMM) API provides full manual control but no automatic bandwidth-aware algorithm, and NVIDIA recommends manually partitioning data using \texttt{cudaMemSet} APIs.
We implement bandwidth-aware page placement with the VMM API, but find that it degrades performance compared to our Linux-API-based design (details in \cref{sec:pam_impl}).

\smallskip
\noindent\textbf{Linux Weighted Interleaving.}
The Linux kernel supports deterministic weighted interleaving of pages across memory tiers via the \texttt{set\_mempolicy()} interface~\cite{mempolicy_linux}, which places pages across
NUMA nodes proportionally to relative tier bandwidth, typically for CPU workloads.
The interleaving is agnostic to the consumer: pages stripe across the address range with no knowledge of which thread---or, on a GPU, which kernel or CTA---consumes which page.
The striping can cross kernel boundaries, so an individual kernel may receive a host share above $\alpha$, underutilizing HBM and incurring slowdown.
Moreover, data a CTA reads over its lifetime need not align to 2MB pages---on the Llama-3.3 down-projection, one page holds 73 of the 128 columns a CTA computes---so the footprint of a CTA spans 1.75 pages, and when one page is host-resident, the host loads stall the SM and delay its co-issued HBM loads.
Finally, striping is fixed at allocation and cannot maintain the $\alpha$ ratio for the KV pages that the serving system provisions and frees at runtime.
As a result, weighted interleaving fails to improve system bandwidth utilization. 
Because no implementation exists for GPU memory, we implement it ourselves and evaluate it in \cref{sec:eval_baselines}.

\subsection{Goal of Our Work}
 
\Cref{tab:properties} compares existing schemes and outlines their key limitations.
No existing scheme provides per-wave concurrency or maintains the $\alpha$ ratio for dynamic KV data.
This work aims to extract the additive HBM+host bandwidth for LLM inference, in addition to scaling capacity.
In our solution, all data accesses from SMs should be demand loads concurrently to both tiers and in proportion to their bandwidth ratio, while performing no data migration at runtime, as prefetching consumes bandwidth.
Ideally, we want to make minimal changes, especially avoiding intrusive modifications to performance-critical kernels in LLM serving systems.

\begin{table}[!htb]
    \centering
    \scriptsize
    \setlength{\tabcolsep}{3pt}

    \caption{Summary of comparisons of current and proposed schemes. All schemes add host capacity but fail to provide per-wave concurrency and do not maintain the $\alpha$ ratio
    for dynamic KV data. $^\dagger$No GPU/LLM implementation exists (\cref{sec:eval}).}
    \label{tab:properties}

    \begin{tabular}{lccccc}
        \toprule
        & \multicolumn{2}{c}{Migration}
        & \multicolumn{2}{c}{Placement}
        & \textbf{Goal} \\
        \cmidrule(lr){2-3}
        \cmidrule(lr){4-5}

        Property
        & \makecell{Prefetching\\\cite{vllm_prefetching}}
        & \makecell{\texttt{cudaMalloc}\\\texttt{Managed}}
        & \makecell{BAPP-R$^\dagger$\\\cite{agarwal2015page}}
        & \makecell{Weighted interleave$^\dagger$\\\cite{mempolicy_linux}}
        & {} \\
        \midrule

        Use host capacity
        & \cmark & \cmark & \cmark & \cmark & \cmark \\

        Migration-free
        & \xmark & \xmark & \cmark & \cmark & \cmark \\

        Proportional accesses
        & \xmark & \xmark & \cmark & \cmark & \cmark \\

        Per-wave concurrency
        & \xmark & \xmark & \xmark & \xmark & \cmark \\

        Works for dyn. data
        & \xmark & \xmark & \xmark & \xmark & \cmark \\

        Key limitation
        & \makecell{HBM\\writes}
        & \makecell{HBM\\thrashing}
        & \makecell{Ratio\\variance}
        & \makecell{Access\\unaware}
        & --- \\

        \bottomrule
    \end{tabular}
\end{table}

\section{Fully Utilizing System Memory Bandwidth}
\label{sec:insights}

\begin{figure}[h]
    \centering
    \includegraphics[width=\linewidth]{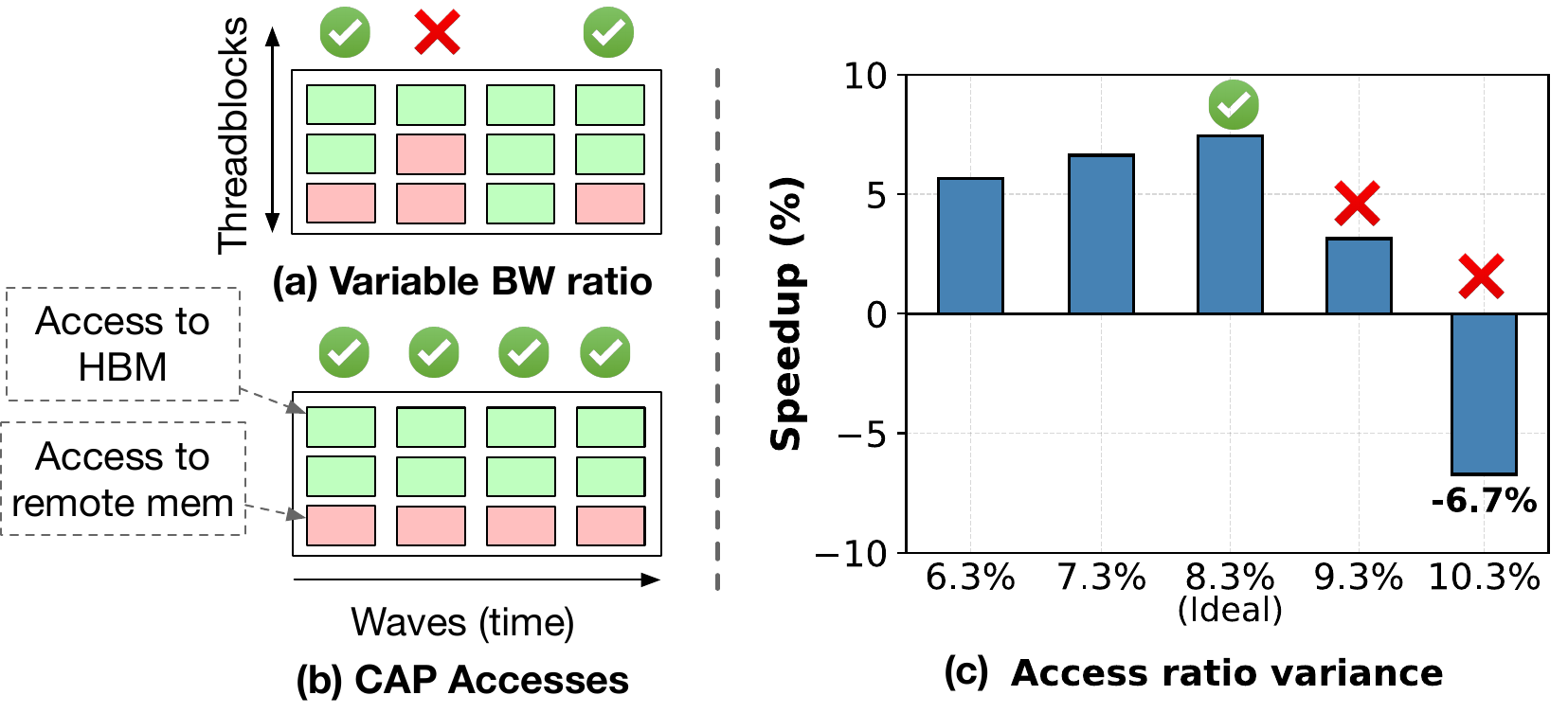}
    \vspace{-4mm}
    \caption{(a) Variable access ratio with BAPP-R. (b) Deterministic access ratio with CAP accesses. (c) Add-kernel speedup at varying access ratios.}
    \label{fig:pf}
    \vspace{-4mm}
\end{figure}

To achieve our goal, we propose \papername, a runtime system that performs concurrent access to host memory and HBM to accelerate LLM serving.
The core idea behind \papername is to leverage access patterns in LLM kernels to guide page placement and runtime data management decisions.
\papername requires addressing three key questions.
\cref{sec:mpp}: Developing a page placement policy that allows for concurrent data access without access ratio variance (\cref{fig:pf}-(a)) or kernel code changes.
\cref{sec:tb_specialize}: Finding the correct granularity of access pattern at the kernel that guides decisions.
\cref{sec:dyn_data}: Handling important LLM kernels (like attention) that access dynamically provisioned data such as KV pairs.

\begin{figure}[h]
    \centering
    \includegraphics[width=0.7\linewidth]{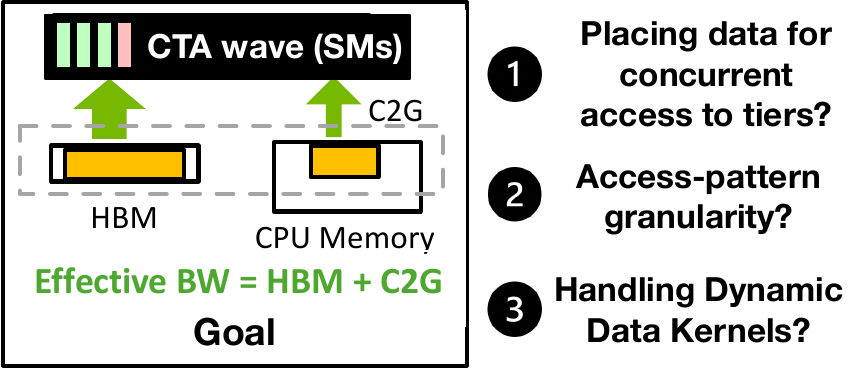}
    \caption{Key questions in enabling concurrent and proportional memory access to fully utilize system bandwidth.}
    \label{fig:challenges}
    \vspace{-3mm}
\end{figure}

\parhead{Concurrent and Proportional Access (CAP).} We introduce the bandwidth-optimal access pattern `CAP' (\cref{fig:pf}-(b)). 
CAP requires partitioning accesses such that \textit{exactly} $\alpha/(1+\alpha)$ of them are to host while $1/(1+\alpha)$ are to HBM to ensure bandwidth proportionality. 
If access ratio to host memory is higher than $\alpha$, we might experience a slowdown. Moreover, CAP requires performing these accesses \textit{concurrently}. Thus, both memory tiers saturate at every wave, eliminating temporal variance to fully utilize system bandwidth.

We first measure the gain that CAP accesses deliver when kernel code is freely editable.
We handcraft the add kernel of \cref{sec:limitations} to perform CAP: the data resides in both tiers, a fixed subset of warps reads from the host tier, and the remaining warps read from HBM (writes stay HBM-local).
\cref{fig:pf}-(c) sweeps the host share of accesses.
The speedup peaks at 7.4\% at a host share of 8.3\%, the measured optimum near the proportional share $\alpha/(1+\alpha)=9.1\%$, and incurs 6.7\% slowdown host ratio considerably exceeds proportional share.
The peak of 7.4\%, against the 10.4\% ideal, is the practical CAP bound for this system because it is difficult for the SMs to saturate both tiers at once (\cref{sec:future}).
Next, we approach this bound for complex LLM serving systems without kernel modifications or data duplication.

\subsection{Wave-aware Modulo Page Placement}
\label{sec:mpp}

We could enable CAP by maintaining two pointers for HBM- and host-resident data and editing the kernel.
We implement this route in a handwritten matmul kernel, which maps host memory with \texttt{cudaHostRegister} and fetches an $\alpha$ fraction of weights from the host tier.
The handwritten kernel gains 7.3\%, 99\% of the CAP bound, so kernel editing extracts nearly the full benefit.
However, two properties of decode kernels make per-kernel editing impractical. The matmul path that dominates dense-decode runtime dispatches to the closed-source cuBLAS library~\cite{cublas}. The open kernels are numerous and specialized: vLLM ships 92 custom CUDA kernels~\cite{vllm}, FlashAttention-3 compiles 401 forward-pass instantiations, and CUTLASS generates its GEMM kernels from templates. A CAP edit recurs across every instantiation and forces re-autotuning after tile-level changes. Prior serving and memory systems adopt transparency as a design goal~\cite{ushare}, as does \papername.
Thus, our \textbf{first challenge} is performing CAP without kernel changes.

Page placement must be aware of the access pattern of each wave, which we term
\textit{wave-awareness}. Expert and MLP kernels use contiguous virtual memory
for model weights and assign each threadblock a virtually contiguous region.
Each wave accesses tens of megabytes, spanning tens of 2MB
pages~(\cref{sec:tb_specialize}). To hold an exact ratio across these few
pages, we use \textbf{modulo-based page placement (MPP)}: for an HBM-to-host
bandwidth ratio of $K=1/\alpha$, MPP allocates the first $K$ pages to HBM and
the $(K{+}1)^{th}$ page to the host tier. For $\alpha=10\%$, MPP matches
$\alpha$ exactly with 11 pages.

\begin{tcolorbox}[boxrule=1pt,left=5pt,right=5pt,top=1.5pt,bottom=1.5pt, colback=green!10]
    \textbf{Insight-1:} \textit{Modulo-based page placement (MPP) maintains the exact bandwidth ratio at every threadblock wave, eliminating the ratio variance of random placement (\Cref{tab:properties}).}
\end{tcolorbox}

\begin{figure}[tb]
    \centering
    \includegraphics[width=0.8\linewidth]{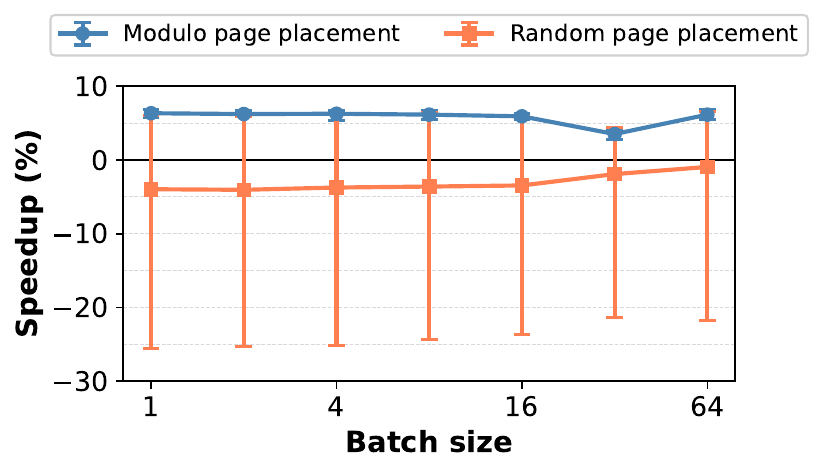}
    \vspace{-3mm}
    \caption{Performance of random and modulo-based page placement for Llama3.3-70B fused gate+up projection matmul kernel (GH200 GPU, details in ~\cref{sec:methodology}) at different batch sizes. Modulo-based page placement consistently outperforms random page placement.}
    \vspace{-3mm}
    \label{fig:random_vs_modulo}
\end{figure}

To analyze MPP, we measure the execution time of the Llama-3.3 70B matmul
kernel (\texttt{fused\_gate\_up} CUTLASS kernel~\cite{cutlass}) across batch
sizes. \cref{fig:random_vs_modulo} plots the average, minimum, and maximum
speedup of MPP and BAPP-R on GH200~(\cref{sec:eval}) over 20 runs, with a
portion of data now in the host tier. MPP outperforms BAPP-R with practically
no variance, whereas BAPP-R varies widely and slows execution on average in
every case (up to 25\%). The average MPP speedup matches the maximum BAPP-R
speedup, so MPP delivers deterministic bandwidth improvement over random
allocation.

MPP is per-kernel, unlike Linux weighted interleaving that stripes data across kernel boundaries, so individual kernels can receive a host share above $\alpha$ (\cref{sec:os_cuda}). 
Thus, BOOST preserves per-kernel proportional access by allocating each weight tensor separately, so the placement pattern restarts at every tensor and kernel's host data stays bounded by $\alpha$ to maximize performance.

\subsection{CTA Specialization for CAP Accesses}
\label{sec:tb_specialize}

Complex GPU kernels set the number of warps and threadblocks (CTAs) at
runtime. Our \textbf{second challenge} is choosing the granularity that guides
page placement: the thread, warp, or CTA level.

\begin{figure}[!htb]
    \centering
    \includegraphics[width=\linewidth]{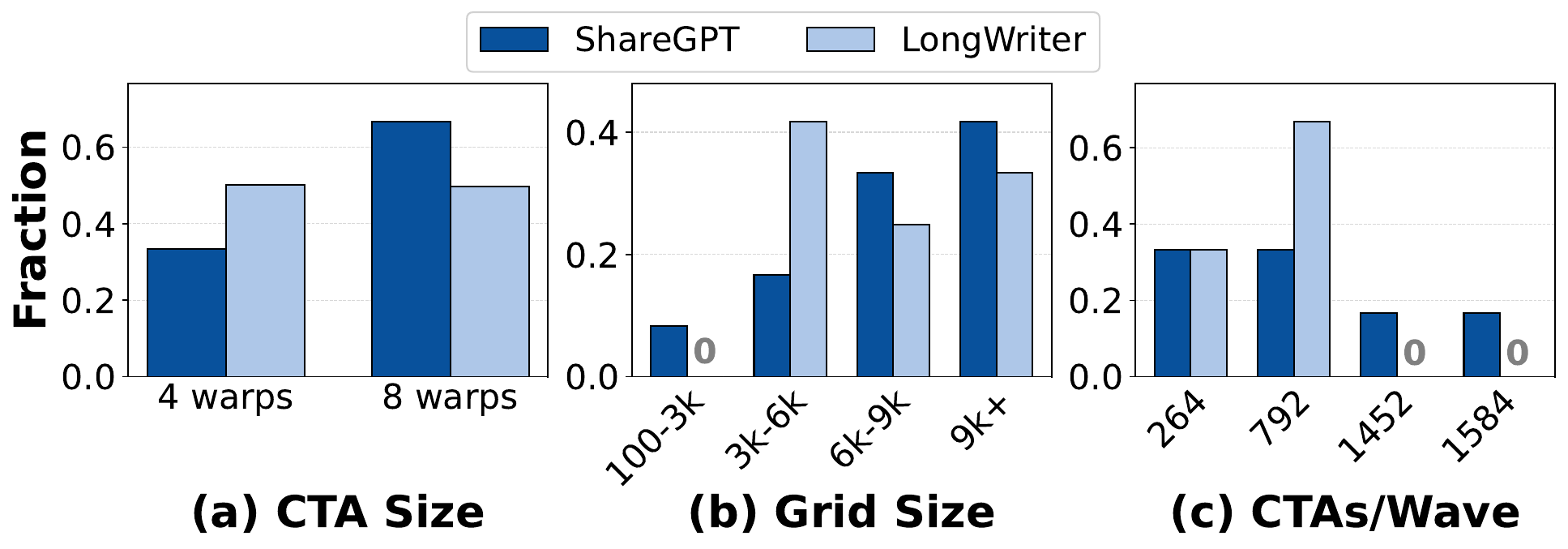}
    \vspace{-3mm}
    \caption{Profiling of Qwen3-Next 80B's expert compute kernel (GH200 GPU) for ShareGPT and LongWriter datasets. Active threads are 264K, CTA size variation is low, but CTAs per wave vary widely depending on problem size.}
    \vspace{-3mm}
    \label{fig:warp_tb_characterize}
\end{figure}

\smallskip
\noindent\textbf{Characterization.} \cref{fig:warp_tb_characterize} plots the
CTA size and grid size of the Qwen3-Next 80B kernel on GH200 across the
ShareGPT~\cite{roschildrui_sharegpt_v3_2023} and
LongWriter~\cite{bai2024longwriter} datasets. The CTA size is 4 or 8 warps,
with low variance, whereas the grid size and active CTAs per wave vary widely.
Every configuration runs 264K active threads per wave (132 SMs $\times$ 2K
threads). Thread- or warp-level granularity might therefore seem best, given
their low size variance.

\begin{figure}[tb]
    \centering
    \includegraphics[width=0.7\linewidth]{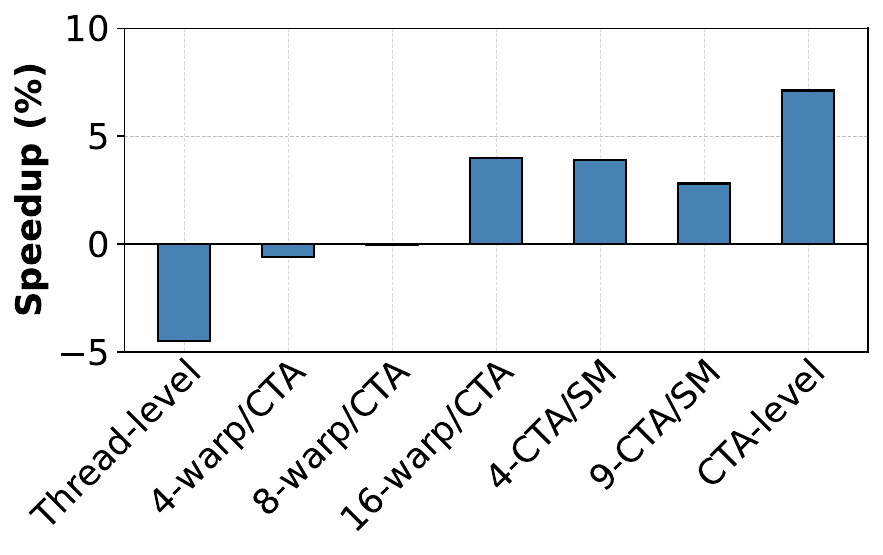}
    \vspace{-3mm}
    \caption{Speedup of Llama-70B up-projection on GH200 compared to HBM-only with different access partitioning.}
    \vspace{-3mm}
    \label{fig:partitioning_speedup}
\end{figure}

\cref{fig:partitioning_speedup} shows the Llama-3.3 70B up-projection performance under each
granularity. Thread-level partitioning introduces warp divergence and prevents
memory-request coalescing. Warp specialization, where warps within a CTA
target different tiers, is too coarse: a 4-warp CTA represents only bandwidth
ratios in multiples of 0.25, and widening the CTA to 16 warps refines control
but degrades the kernel. Confining a subset of SMs to host accesses
(`4-CTA/SM') lowers memory-level parallelism and underutilizes HBM. CTA
specialization avoids these failures: hundreds to thousands of CTAs run per
wave~(\cref{fig:warp_tb_characterize}), giving fine control, and CTAs execute
independently on SMs without synchronization overhead.

Concurrency requires that we confine host accesses to a subset of CTAs. When the accesses of a single CTA
span both tiers, the host loads stall the SM and delay the co-issued HBM loads, so the SM loses HBM
bandwidth.
The slowdown is shown with thread-level partitioning in \cref{fig:partitioning_speedup}.
Restricting host accesses to an $\alpha$ fraction of CTAs instead lets the remaining CTAs
saturate HBM and overlap with the latency-bound CTAs.

\begin{tcolorbox}[boxrule=1pt,left=5pt,right=5pt,top=1.5pt,bottom=1.5pt, colback=green!10]
    \textbf{Insight-2:} \textit{Guiding page placement on memory tiers using CTA wave accesses facilitates CAP for typical kernels.}
\end{tcolorbox}
 
This insight also keeps each CTA on a single tier for its lifetime: an
$\alpha/(1+\alpha)$ fraction of the CTAs fetch from the host tier exclusively.
CTA-level placement gains 7.1\% (\cref{fig:partitioning_speedup}), 96\% of the 7.4\% CAP bound compared to 99\% for the handwritten kernel (\cref{sec:mpp}).
Placement alone recovers the gain of kernel editing, obviating kernel changes.

Even with CTA-level placement, the footprint of a CTA need not align to page boundaries: on the Llama-70B down-projection, the footprint of one CTA spans 1.75 pages (\cref{sec:os_cuda}), so under MPP one of these pages could land on the host tier which splits CTA-level accesses, incurring slowdown.
For such kernels, \papername allocates at the granularity of \emph{superpages}: contiguous page groups sized to the least common multiple of the CTA footprint and the page size, so CTA footprint does not span a superpage boundary.
\papername places each superpage in one tier and applies MPP across superpages, limiting CTA-level accesses to one memory tier (\cref{sec:design_impl}).

\subsection{Wave-aware Page-to-Data Mappings}
\label{sec:dyn_data}

Modulo page placement works for expert and MLP kernels because their
threadblocks access contiguous, statically-allocated weights. Enabling CAP for attention is difficult because serving systems use paged attention~\cite{pagedattention_vllm}, which
inserts a free pool of KV pages between virtual memory and threadblock data.
As requests arrive and complete, the pool provisions and frees KV pages, and
the churn makes the mapping from virtual pages to the active wave near-random
over time, even under MPP.
Our \textbf{third challenge} is maintaining CAP
despite this dynamic mapping.

The problem is that the placement is fixed at allocation, while the free pool reassigns pages to data at runtime (\cref{sec:os_cuda}).
Static allocation-time policies (MPP, BAPP-R, Weighted Interleaving) are unable to dictate a page-to-data mapping that changes dynamically.
The pages a wave touches are decided by the allocation order of the free pool, so an interleave that is proportional over the address range still loses the per-wave ratio once blocks churn.

CAP for dynamic data therefore needs awareness of the data structures the kernel accesses, beyond kernel- and CTA-awareness.
Placing an $\alpha$ fraction of whole requests on the host tier would satisfy
proportionality, but unknown context lengths could let long requests land on the
host tier and raise tail latency, so we partition at a finer granularity.
We observe that recent attention kernels~\cite{pagedattention_vllm, ye2025flashinfer,
shah2024flashattention} assign each KV head to different threadblocks and
chunk the context of a head across further threadblocks, so the CTAs in a wave
process different \textit{KV head chunks}.
\begin{tcolorbox}[boxrule=1pt,left=5pt,right=5pt,top=1.5pt,bottom=1.5pt,colback=green!10]
    \textbf{Insight-3:} \textit{Attention kernel threadblocks operate on `KV head chunks', and mapping $\alpha$ fraction of chunks to pages in host tier at runtime facilitates CAP accesses.}
\end{tcolorbox}

\begin{figure}[t]
    \centering
    \includegraphics[width=0.7\linewidth]{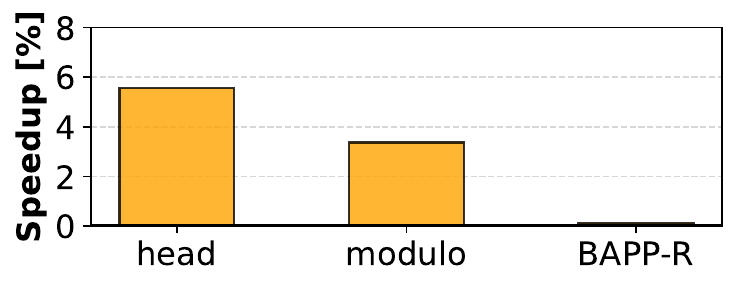}
    \vspace{-3mm}
    \caption{Speedup of FlashAttention3~\cite{shah2024flashattention} kernel for Llama3.1-8B on GH200 for different mappings. Managing page-to-KV head chunk mapping maximizes TPOT speedup.}
    \label{fig:chunk_speedup}
    \vspace{-3mm}
\end{figure}

\begin{figure}[!htb]
    \centering
    \includegraphics[width=0.95\linewidth]{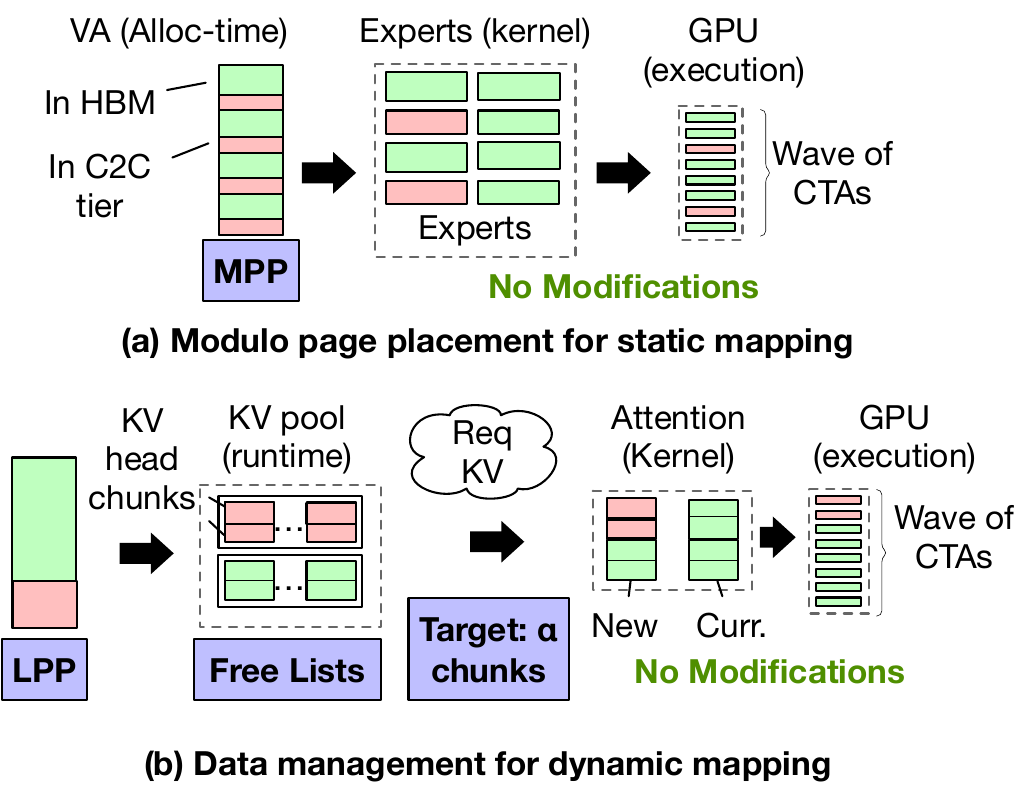}
    \vspace{-3mm}
    \caption{Top: MPP ensures CAP with static data (model weights). Bottom: Wave-aware LLM system for CAP with dynamic KV data. LPP is linear (BW-proportional) placement.}
    \label{fig:static_dynamic}
    \vspace{-3mm}
\end{figure}

Dynamic-data kernels need
two levels: linear placement for average proportionality, and a separate \textit{runtime
fulfillment mechanism} that maintains the per-wave ratio under churn---which static
interleaving techniques cannot provide (the dynamic-data row of \cref{tab:properties}).
The runtime fulfillment maps chunks greedily to keep an $\alpha$ fraction of active chunks on the
host tier at runtime. As \cref{fig:chunk_speedup} shows, this chunk-aware
policy outperforms MPP and BAPP-R, yielding end-to-end TPOT speedup even when
only the attention kernel performs CAP.
Head-chunk mapping gains 5.9\% on FlashAttention-3 (81\% of the CAP bound), and the fulfillment holds the host fraction within $\pm$1\% of the ideal ratio at runtime (\cref{sec:eval}).
We address practical challenges
(fragmentation, tensor reshaping) in \cref{sec:dsr_impl}.
\cref{fig:static_dynamic} summarizes the key insights of our design.

\section{\papername Design and Implementation}
\label{sec:design_impl}

\papername enables CAP accesses on every wave of LLM serving without kernel changes.
It places data across both HBM and host memory, so it increases the usable capacity over HBM-only serving in addition to raising bandwidth utilization.
Every access is a cache-coherent demand load to one tier, and \papername performs no data migration at runtime and is simple to implement.

\subsection{Design Overview}
\begin{figure}[h!]
    \centering
    \includegraphics[width=0.67\linewidth]{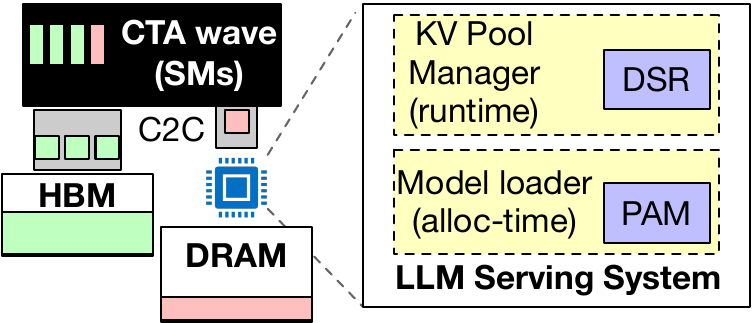}
    \vspace{-3mm}
    \caption{Design overview of \papername.}    \label{fig:boost_design_overview}
    \vspace{-5mm}
\end{figure}

\cref{fig:boost_design_overview} shows the two components of \papername.
The Page Allocation Manager (PAM) places pages at allocation time and realizes the wave-aware placement of \cref{sec:mpp} and \cref{sec:tb_specialize}.
The Data Shape Remapper (DSR) manages dynamic KV data at runtime and realizes the runtime fulfillment of \cref{sec:dyn_data}.
Prior heterogeneous-memory work~\cite{agarwal2015page} and CPU-NUMA systems~\cite{mempolicy_linux} place pages bandwidth-proportionally at the memory allocator and the device driver.
These layers observe only the address space.
CAP additionally requires awareness of kernel access patterns and runtime management of dynamic data structures~(\cref{sec:insights}), and only the serving system has this information, because it loads the model weights and manages the KV pool.
\papername therefore implements both components inside the serving system.
Both components run on the CPU, and \papername adds no GPU code.

\subsection{Page Allocation Manager (PAM) Design}
PAM resides in the allocation path of the serving system and places pages according to the data each kernel accesses.
For statically allocated data, such as the model weights of expert and MLP kernels, PAM applies modulo page placement (MPP) across the two tiers, and uses superpages for kernels whose CTA footprint misaligns with pages~(\cref{sec:tb_specialize}).
For dynamically mapped data, such as the KV pairs of the attention kernel, allocation-time placement alone cannot hold the per-wave ratio~(\cref{sec:dyn_data}), so PAM provides the first of the two management levels.
PAM applies linear page placement (LPP), which partitions the virtual address space contiguously: the first $1/(\alpha+1)$ of pages map to HBM and the last $\alpha/(\alpha+1)$ map to the host tier~(\cref{fig:static_dynamic}-bottom left).
Because LPP makes the tier of a page a fixed function of its virtual address, DSR controls the physical placement of a KV block by choosing its virtual address.
PAM overrides the default allocation routines of the serving system, enabling kernels to perform CAP accesses transparently.

\subsection{Data Shape Remapper (DSR) Design}
DSR provides the second management level for dynamic data. DSR extends the runtime of the serving system so that the per-wave ratio is maintained approximately rather than exactly.
The KV block manager of the serving system requests and frees \textit{blocks} of KV pairs, and each block holds a fixed number of pairs (often 16).
DSR maintains a bandwidth-proportional access ratio at every wave through two mechanisms.
First, DSR maps the KV head chunks within each block to the two tiers through LPP-allocated virtual memory.
Second, DSR fulfills block requests so that an $\alpha$ fraction of the active KV head chunks stays on the host tier.

\parhead{Block-level KV Head Chunk Partitioning.}
DSR partitions KV data along two axes---attention heads and requests.
Kernels assign different KV heads to different threadblocks, but many LLMs expose too few heads for fine control.
Qwen3-30B, for example, has four KV heads, and placing one head on the host tier already sends 25\% of the KV accesses there, far above the $\alpha=10\%$ target.
DSR therefore defines two block types.
A \textit{single-head-host} block maps one head to the host tier and the remaining heads to HBM.
An \textit{all-local} block maps all heads to HBM.
For a target of $1/M$ of the KV data on the host tier with $H$ heads, DSR maintains single-head-host and all-local blocks in the ratio $H$ to $M-H$, so the host-resident share of KV data averages $1/M$ over time.

\parhead{Wave-aware Block Fulfillment.}
As requests arrive and complete, the allocator tracks the distribution of free blocks to hold the target ratio.
When a new request needs KV cache, DSR assigns the single-head-host type if the free single-head-host fraction exceeds $H/M$, and assigns the all-local type otherwise.
This greedy policy keeps the host-fetched fraction of KV head chunks near $\alpha$ across request lengths and arrival patterns.
All blocks of a request share one type, which reduces temporal variance across waves.

The greedy policy meets two boundary conditions when a free list empties.
When the single-head-host list empties, DSR assigns the all-local type and the host fraction falls below $\alpha$.
The system forfeits part of the bandwidth benefit but incurs no slowdown, because only a fraction above $\alpha$ oversubscribes the host tier~(\cref{sec:insights}).
When the all-local list empties, DSR examines the number of free single-head-host blocks.
Typically, the shortage stems from an imbalance in request lengths, and DSR appends single-head-host blocks to a request of the all-local type.
The assignment is sticky, so the request continues to receive single-head-host blocks and the mixing of block types stays confined to that request.
When total available blocks are low and the system faces critical memory pressure, DSR defers to the scheduler of the serving system.

\subsection{PAM Implementation Details}
\label{sec:pam_impl}

\parhead{Superpages.}
MPP at page granularity assumes the weight footprint of each CTA aligns to a page boundary.
In practice, a CTA computes an output tile of 64--256 columns, and each column spans the full contraction dimension of the weight.
The contraction dimension of some kernels is a non-power-of-2 value, such as 28K on the Llama down-projection, so each FP8 column occupies 28KB and the footprint of one CTA reaches 3.5MB, which spans page boundaries~(\cref{sec:tb_specialize}).
For such kernels, PAM sizes superpages to the least common multiple of the per-CTA footprint and the page size (e.g., LCM(3.5MB, 2MB) = 14MB for Llama-70B down-projection), places each superpage in one tier, and applies MPP across superpages to bound the host fraction to $\alpha$ relative to HBM.

\parhead{Interleaving on Cache-Coherent Systems.}
We implement PAM in vLLM's \texttt{model\_loader}. Grace Hopper is cache-coherent, so the OS exposes GPU
memory as a NUMA node and standard Linux APIs interleave CPU and GPU memory in one contiguous virtual
range. For each 2MB region (\texttt{MAP\_HUGETLB}, over the 64KB default, to limit TLB thrashing), PAM
calls \texttt{mmap} with \texttt{mbind} to select the node and \texttt{cudaHostRegister} to expose it to
the GPU, then zeros it to force physical backing.
For MPP, PAM maps $K$ superpages to HBM and one superpage to the host tier in turn, and LPP allocations are
similarly bandwidth-proportional. When $K=1/\alpha$ is not an integer, PAM sets $K = \lceil 1/\alpha
\rceil$, which keeps the host fraction below $\alpha$; a fraction above $\alpha$ oversubscribes the
host tier and incurs slowdown~(\cref{sec:insights}).
A small residual variance remains, because the final wave of a kernel holds fewer CTAs than the
rest. However, eliminating this variance requires the wave count in advance, which are not currently exposed by parameterized CUTLASS kernels.

\subsection{DSR Implementation Details}
\label{sec:dsr_impl}

\parhead{Virtual Memory Integration.}
DSR extends the KV block manager to maintain two free lists, one for each block type.
The two lists exploit the existing virtual memory indirection of the serving system to control physical placement, so DSR requires no per-page tracking and no changes to the allocation logic of the kernels.

\parhead{Tensor Layout Reordering.}
A single KV block, holding 16 tokens across all layers and heads, occupies approximately 2.5MB in models like Llama-3.3 70B-FP8.
The block size misaligns with the 2MB page size, so the heads of one block straddle page boundaries and DSR cannot place a single head on the host tier directly.
DSR resolves the misalignment by reordering the tensor layout.
The paged attention kernels of vLLM describe tensors using base pointers, dimension lengths, and strides, which permits non-contiguous virtual memory as long as the access pattern stays regular.
DSR reshapes the KV tensor to place heads as the outermost dimension, so the data of each head becomes contiguous and aligns to the LPP partition.
The reordered layout preserves the locality optimizations of the kernel within each head and bounds the host access ratio by $1/H$ even under maximum imbalance.
When $H > M$, DSR mixes blocks with $\lfloor H/M \rfloor$ and $\lceil H/M \rceil$ host heads to maintain the target ratio.

\begin{figure*}[t!]
    \centering
     \includegraphics[width=0.9\textwidth]{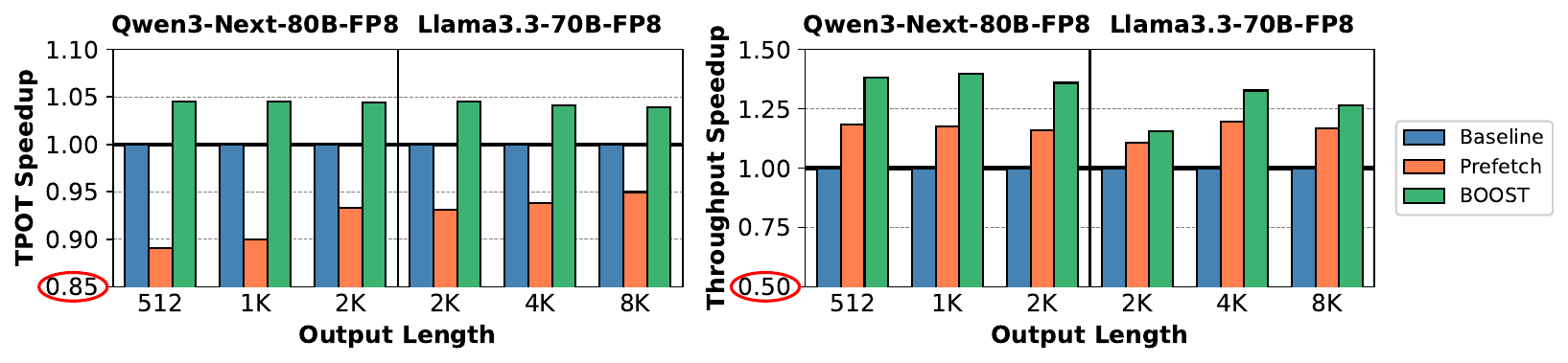}    
     \vspace{-3mm}
     \caption{Left: TPOT speedup for Llama3.3 dense LLM (batch size of 10) and Qwen3-Next MoE (batch size of 88) for low latency serving, normalized to HBM-only baseline. \papername improves TPOT by 4.3\% on average (geometric mean) while prefetching degrades TPOT by 6\% for iso-batch serving. Right: Normalized throughput speedup in high throughput serving, where batch size scales to fill the available memory capacity. \papername provides 31\% average throughput speedup, 15\% higher than prefetching.} 
    \label{fig:main_combined}
\end{figure*}

\section{Evaluation Results}
\label{sec:eval}

\subsection{Methodology}
\label{sec:methodology}
 
\parhead{Serving System.} We extend vLLM (v0.17.0) and our changes are
approximately 800 LoC (Python and PyTorch).
 
\parhead{Models.} We evaluate two recent LLMs for our main evaluations:
Qwen3-Next-80B MoE~\cite{qwen3technicalreport} and Llama3.3-70B~\cite{grattafiori2024llama} dense LLM. We
also evaluate the larger Qwen2.5-72B~\cite{qwen2025qwen25technicalreport}, for scalability study with
tensor parallelism. Unless specified, all models are in FP8.
 
\parhead{Workloads.} We use synthetic workloads with varying context
lengths. For the dense model we use an input/output length of
256/2K, 512/4K, 1K/8K. For the MoE model we use a smaller input/output length
of 64/512, 128/1K, 256/2K. We also show sensitivity to increasing context
length (\cref{sec:ctx_sens}) and batch sizes (\cref{sec:batch_sens}).

\parhead{Experimental Setup.}
We use a Grace Hopper system (GH200) with 96GB HBM and 480GB CPU memory. Peak GPU HBM bandwidth is 3.63TiB/s and CPU memory bandwidth is 476GiB/s. GPU connects to the CPU memory tier via cache-coherent NVLink C2C at 419GiB/s. We measure peak HBM and C2G bandwidth at 3330GiB/s and 350GiB/s, respectively ($\alpha$=$10\%$). All main evaluations use the Grace Hopper system. We evaluate systems with higher $\alpha$ of 12\% using H100-based emulation of the host tier (details in \cref{sec:bw_ratio_sens}).
 
\parhead{Serving Scenarios and Metrics.}
All experiments use prefill-decode disaggregation, the standard production
configuration~\cite{sglang_gb200, sglang_pd}.
Our GH200 node has a single GPU, so the prefill and decode instances
time-share the GPU, and we report the metrics of the decode instance.
We evaluate all systems under two representative LLM serving scenarios (\cref{sec:cap_bw_isolation} provides the performance decomposition):
 
\begin{itemize}
    \item \textbf{Low-Latency Serving.} The goal is to minimize time-per-output-token (TPOT). We fix the maximum runnable batch
    size (batch of 88 requests for Qwen3-Next and 10 requests
    for Llama-70B) and measure TPOT as the average time to
    generate each output token during the decode phase.
    The batch size is fixed, so any TPOT change stems solely from the additional system bandwidth utilization.
 
    \item \textbf{High-throughput Serving.} The goal is to maximize tokens per second (TPS) throughput. The serving system
    automatically adjusts the batch size to fully utilize the available memory capacity. We measure TPS as the total number
    of tokens generated (including the first token) divided by
    the total execution time.
    The auto-scaled batch lets the added capacity of the host tier raise concurrency, so TPS captures the capacity and bandwidth benefits together.
\end{itemize}

\parhead{Evaluated Policies.}
\begin{itemize}
    \item \textbf{Baseline.} The baseline system allocates all data in HBM
    without using C2G memory. The setup represents standard
    LLM serving when the model fits in the GPUs and does not
    utilize the additional host tier memory bandwidth.
 
    \item \textbf{Prefetch.} Asynchronous prefetching
    in vLLM~\cite{vllm_prefetching} stores one out of every K layers in host memory (1/$K$
    approximates $1/(1 + \alpha)$) and asynchronously prefetches it
    to HBM before use (we use $K = 12$ to maximize speedup).
 
    \item \textbf{\papername.} We evaluate our solution that employs MPP for
    static model weights and LPP with runtime $\alpha$ management
    for dynamic KV data.
    The placement modulus is 12: one of every 12 pages resides on the host tier, a host share of 8.3\% that matches the measured-optimal share of \cref{sec:insights}.
\end{itemize}

We additionally evaluate CUDA's driver-based migration policy, \texttt{cudaMallocManaged}.
Finally, we evaluate BAPP-R and Linuxed weighted interleaving policies, representing
SoTA bandwidth-aware page placement~\cite{agarwal2015page, chou2017batman, mempolicy_linux}.

\subsection{Key Results: TPOT and Throughput Speedup}
\cref{fig:main_combined} plots normalized TPOT for low-latency serving (left) and system throughput for
high-throughput serving (right) across Qwen3-Next and Llama-3.3 at varying context lengths, normalized to
the HBM-only baseline. At the same batch size, \papername improves TPOT by 4.3\% on average. Prefetching
degrades TPOT by 6\%, making \papername 11\% faster.
\papername provides consistent TPOT benefit for both dense and MoE models. Prefetching, however, suffers more
for the MoE: dynamic expert activation patterns introduce variability in layer execution time, causing
prefetches to miss their timing window. This leads to a 9.3\% average TPOT degradation for the MoE. \papername
is unaffected because it performs cacheline-granular demand access to both tiers concurrently, requiring
no orchestration of data movement, which has the added benefit of reducing implementation complexity and finetuning requirements.
 
\papername improves system throughput by 31\% on average, and up to 40\%, as the 10\% additional capacity from
the host tier substantially increases the runnable batch size.
The added capacity drives most of
this throughput gain; \cref{sec:cap_bw_isolation} isolates the bandwidth-concurrency component at about
4\%. Prefetching improves throughput by 17\% on average but suffers from data migrations that consume HBM
read bandwidth.
Throughput improvement of Qwen3-Next 80B is higher because it is more capacity-constrained at 80GB out of
96GB taken by model weights, so only about 16GB is available for KV cache. By improving effective GPU
capacity by 10\%, or about 9.6GB, \papername provides 60\% additional space for KV, while also increasing
system bandwidth.
\papername is functional for all vLLM kernels that access model weights or KV pairs, including
FlashAttention-3~\cite{shah2024flashattention} and closed-source cuBLAS routines~\cite{cublas} used by MLP
kernels.

\ignore{
- mention TPOT and TPS as key metrics
- two scenarios: low-latency (iso batch size) and high-throughput (max batch size)
- low-latency: no prefill chunking, TPOT measurement, online serving
- throughput-oriented: prefill chunking, TPS measurement (including first token), offline serving
}

\subsection{Memory Bandwidth Utilization}
\begin{figure}[htb!]
    \centering
    \includegraphics[width=0.95\linewidth]{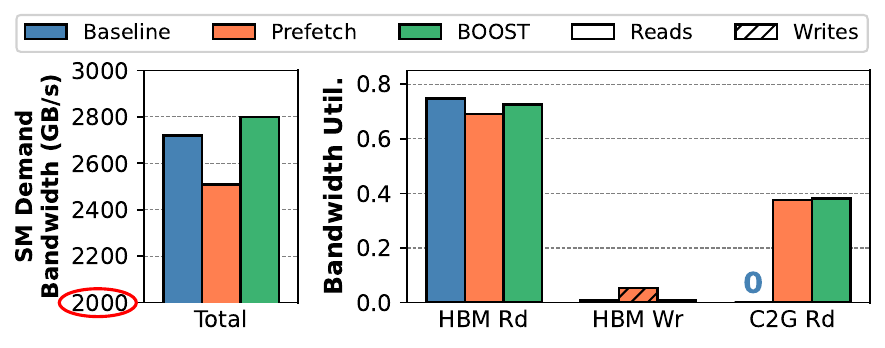}
    \vspace{-3mm}
    \caption{Bandwidth utilization (left) and its breakdown (right) of HBM and C2G for Llama-3.3 (512/4K context). \papername improves bandwidth utilization (HBM+C2G) by 3\% over HBM-only baseline and by 12\% over prefetching.}    \label{fig:main_bw}
    \vspace{-3mm}
\end{figure}

We measure the average bandwidth utilization in the low-latency serving setup with the 4K-context workload. Using Nvidia Nsight Systems~\cite{nsight}, we measure average utilization of the decode phase of a request. \cref{fig:main_bw} shows the total bandwidth utilization (left) and normalized HBM and C2G bandwidth utilization for various policies. 
Prefetching degrades the demand bandwidth utilization available to the SMs by 7.8\% compared to HBM-only serving, as writes due to data migration contend with demand reads over the simplex HBM interface.
\papername improves the system bandwidth utilization by 3\% over HBM-only serving and by 12\% over prefetching.

\subsection{Performance Contribution Breakdown}
 
We isolate the contributions of PAM and DSR using system throughput: PAM alone on
the expert and MLP kernels (\papername-MLP-only) versus the full system (\papername), which
adds DSR to the attention kernel. Table~\ref{tab:perf_breakdown} reports
throughput for Llama-3.3 70B, normalized to HBM-only at the maximum runnable
batch of 10. The expert and MLP kernels hold most of the decode time in large
models, so PAM alone delivers 1.23x. Adding DSR for attention yields a further
7.2\% (1.32x total).
DSR contributes less than PAM because attention occupies roughly
30\% of decode runtime (Nsight); model weights dominate, and PAM serves them.
The contribution of DSR---9 of the 32 points in Table~\ref{tab:perf_breakdown}, or 28\%
of the total gain---is bounded near the runtime share of attention. This share
grows at long context, where the attention footprint grows relative to the
weights.
Overall, the two optimizations are additive.

\begin{table}[!htb]
\centering
\small
\caption{Performance of \papername for Llama-3.3 70B (512/4K in/out context) with PAM (\papername-MLP-only) and with both PAM+DSR (\papername). Our optimizations are additive.}
\label{tab:perf_breakdown}
\begin{tabular}{|l|l|l|l|}
\hline
\textbf{Model/ Config} & \textbf{HBM-only} & \textbf{\papername (MLP-only)} & \textbf{\papername} \\ \hline
Llama-3.3 70B          & 1026          & 1263 (1.23x)              & 1352 (1.32x)              \\ \hline
\end{tabular}
\end{table}

\subsection{Isolation of Capacity and Bandwidth Gains}
\label{sec:cap_bw_isolation}
 
\papername raises throughput two ways: the host tier enlarges the runnable batch,
and concurrent access to both tiers raises demand bandwidth. Isolating them
requires an HBM-only baseline with the same total capacity as \papername. Lacking a
GH200 above 96GB, we run \papername below full HBM utilization and grant the
HBM-only baseline equal total capacity entirely in
HBM~(\cref{tab:cap_bw_isolation}). For Llama-3.3 70B at 1K/8K context, this
capacity-matched baseline raises throughput by 55\% over the HBM-only
reference, a capacity component available to any unified-memory mechanism. At
equal capacity and batch, \papername improves both throughput and TPOT by a further
4\%. This bandwidth-concurrency component is unique to \papername; no capacity-only
mechanism provides it~(\cref{tab:baselines}), and it is
consistent with the 4.3\%
iso-batch TPOT gain of \cref{fig:main_combined}.

\begin{table}[h]
\centering
\scriptsize
\caption{\papername speedup decomposition for Llama-3.3 70B (1K/8K). The capacity-matched baseline holds
the same capacity as \papername in HBM, isolating bandwidth-induced gains.}
\label{tab:cap_bw_isolation}
\begin{tabular}{lccc}
\toprule
\textbf{Configuration} & \textbf{Capacity (BS)} & \textbf{Throughput (TPS)} & \textbf{TPOT (ms)} \\
\midrule
HBM-only (85\% HBM mem-util)             & 81.6GB (BS=6) & 222 & 27.1 \\
HBM-capacity (95\% HBM)      & 91.2GB (BS=10) & 344 (1.55x) & 29.0 \\
\papername (85\% HBM + CPU-mem)    & 91.2GB (BS=10) & 359 (1.62x) & 27.9 \\
\bottomrule
\end{tabular}
\end{table}

\subsection{Comparison: BAPP-R}
 
\begin{figure}[htb!]
    \centering
    \includegraphics[width=0.9\linewidth]{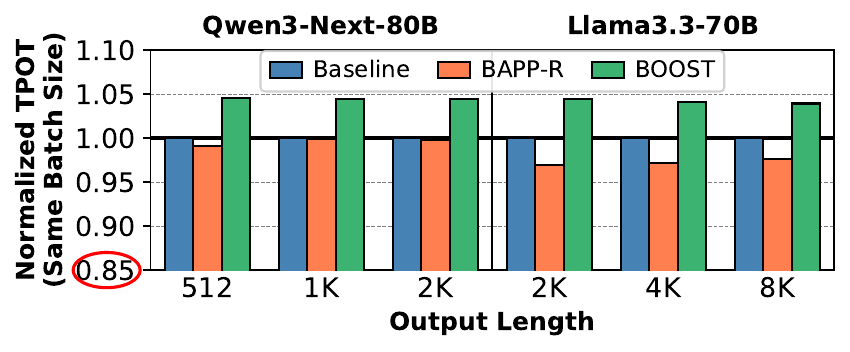}
    \caption{TPOT speedup of \papername and BAPP-R for Qwen3-Next and Llama-3.3. BAPP-R provides 0.4\% and 2.7\% TPOT degradation for MoE and dense LLM, respectively.}    
    \label{fig:bappr_comparison}
    \vspace{-4mm}
\end{figure}

We implement BAPP-R by performing bandwidth-proportional and random page allocation for both model weights and KV pairs.
As \cref{fig:bappr_comparison} shows for Llama-3.3 and Qwen3-Next models for low-latency setting, BAPP-R  causes a 0.4\% and 2.7\% TPOT degradation for the MoE and dense LLM, respectively.
In contrast, \papername provides consistent TPOT gains of 4.5\% and 4.2\% for the MoE and dense LLM. 
Random placement suffers from the ratio variance of \cref{tab:properties}---the access ratio to the two tiers varies across waves, which degrades bandwidth utilization.
We measure the average memory bandwidth utilization for BAPP-R using Nsight Systems (not plotted for brevity) and find that it degrades demand bandwidth to the SMs by 3.6\% compared to HBM-only serving and by 6.4\% compared to \papername.
The lower bandwidth utilization results in worse TPOT for BAPP-R.
 
\revhl{

}

\subsection{Comparison: OS and CUDA Schemes}
\label{sec:eval_baselines}
 
We evaluate OS and CUDA capacity-expansion schemes in \cref{tab:baselines}.
Weighted interleave provides proportionality and raises throughput by 30\%,
but lacks concurrency and fails to improve iso-batch TPOT (0.99), while \papername
delivers 35\% throughput and 4\% TPOT speedup at the 512/4K configuration of \cref{tab:baselines}.
Weighted interleave splits the footprint of a CTA across tiers and stripes across kernel boundaries (\cref{sec:limitations}). BOOST resolves these problems with superpages and per-tensor MPP (\cref{sec:tb_specialize}, \cref{sec:mpp}).
\papername confines the accesses of each CTA to one tier
through MPP and superpages~(\cref{sec:design_impl}) and bounds the host fraction of accesses by every
kernel to at most $\alpha$.
 
Proportional interleave maps contiguous pages to each tier, so the accesses of
early waves target HBM and those of late waves target the host
tier\revarxiv{;}
this reduces to on-demand fetching and drops
performance to 0.6x on the up-projection, a 67\% slowdown. cudaMallocManaged
migrates pages toward the accessing processor: it matches HBM-only when the
working set fits in HBM (1.00), but once the set exceeds HBM the migrations
thrash HBM as a page cache, and each access reduces to an on-demand host fetch
(throughput of 0.19x, a 5$\times$ slowdown). \papername performs no migration; it pins
placement through \texttt{mbind} and lets the SM issue 64B cache-coherent
demand loads to both tiers.

\begin{table}[htb!]
\centering
\scriptsize
\caption{Comparison of capacity expansion techniques using Llama3.3-70B (512/4K). Existing techniques fail to deliver consistent speedup (per-row maximum in bold). All four techniques (columns) use identical batch sizes.}
\label{tab:baselines}
\begin{tabular}{|c|c|c|c|c|}
\hline
\textbf{Llama Norm. perf} & \textbf{\begin{tabular}[c]{@{}c@{}}Weighted \\ interleave\end{tabular}} & \textbf{\begin{tabular}[c]{@{}c@{}}Proportional\\ interleave\end{tabular}} & \textbf{cudaMallocManaged} & \textbf{\papername} \\ \hline
up-proj kernel             & 1.06                                                                    & 0.60                                                                       & 1.00                       & \textbf{1.07}  \\ \hline
down-proj kernel           & 0.99                                                                    & 1.00                                                                       & 1.00                       & \textbf{1.03}  \\ \hline
Iso-batch TPOT      & 0.99                                                                    & 0.75                                                                       & 1.00                       & \textbf{1.04}  \\ \hline
Max throughput      & 1.30                                                                    & 1.08                                                                       & 0.19                       & \textbf{1.35}  \\ \hline
\end{tabular}
\end{table}

\subsection{Impact on Tail Latency}
\label{sec:tail_latency}
 
\begin{figure}[htb!]
    \centering
    \includegraphics[width=0.75\linewidth]{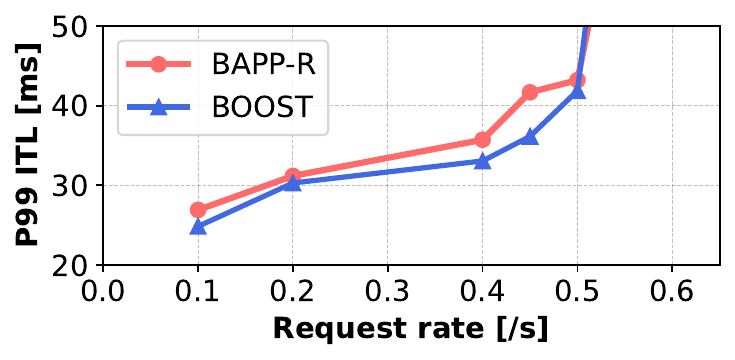}
    \caption{P99 ITL versus request rate for \papername and BAPP-R on the decode instance (Llama-3.3 70B,
    LongWriter). \papername reduces P99 ITL by 7.8\% (geometric mean) over BAPP-R across the serviceable
    load range.}
\label{fig:p99_itl}
    \vspace{-3mm}
\end{figure}

\cref{fig:p99_itl} plots the P99 inter-token latency (ITL), the 99th percentile of the per-token latency during decode, against request rate for \papername and BAPP-R (Llama-3.3 70B,
LongWriter). Requests arrive with Poisson inter-arrival times.
\papername holds a lower P99 ITL than BAPP-R at every serviceable request rate, by 3--13\% (7.8\% geometric mean),
reflecting its higher demand bandwidth. Both systems share the same host-tier capacity and saturate near
0.55 req/s, so under a fixed P99 ITL service level agreement, \papername sustains a higher request rate. The primary advantage of \papername over BAPP-R is lower per-token latency at each request rate.

\subsection{Context Length Variation}
\label{sec:ctx_sens}
\begin{figure}[htb!]
    \centering
    \includegraphics[width=0.95\linewidth]{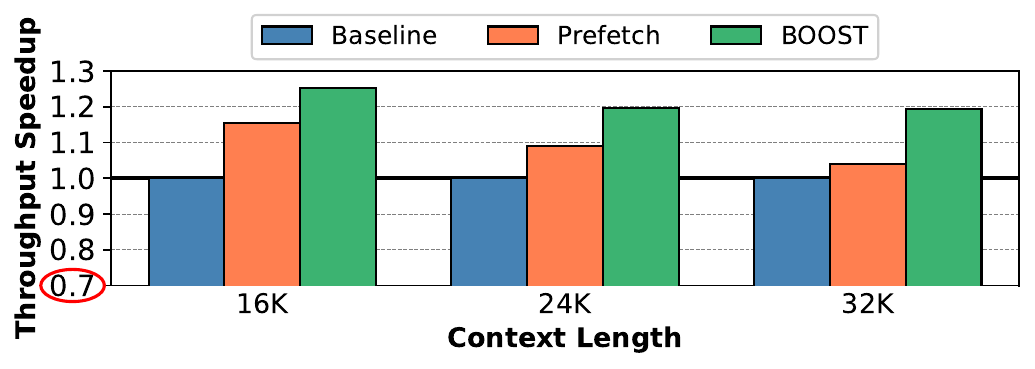}
    \vspace{-4mm}
    \caption{Normalized throughput speedup of Llama3.3 model with longer context (output fixed at 8K). \papername scales to long context workloads.}    \label{fig:main_long}   
    \vspace{-3mm}
\end{figure}

 We evaluate Llama-3.3 model with longer output context lengths and plot throughput speedup in \cref{fig:main_long} normalized to HBM-only serving. We fix the (synthetic) input context to 1K. 
 \papername consistently outperforms prefetching across all context lengths by 11\% on average.
 \papername's speedup drops slightly from 25\% at 16K context to 19\% at 32K context.
 Overall, \papername scales to longer context workloads, making it viable for reasoning and agentic systems.
 
\subsection{Sensitivity: Batch Size}
\label{sec:batch_sens}
 
\begin{figure}[htb!]
    \centering
    \includegraphics[width=0.9\linewidth]{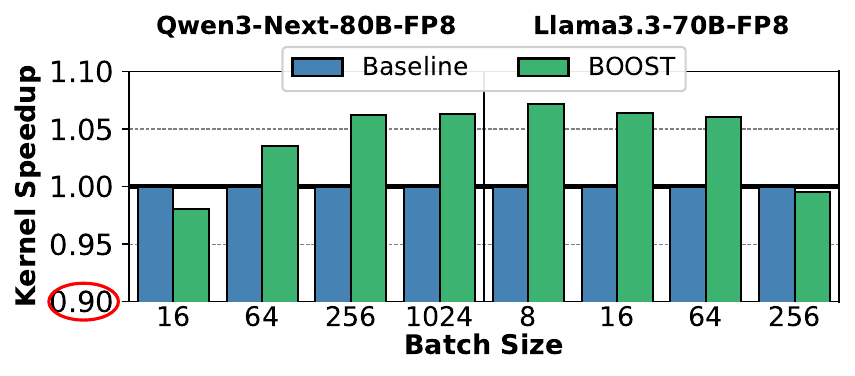}
    \vspace{-4mm}
    \caption{Speedup of expert kernel from a MoE model (Qwen3-Next) and matmul kernel from a dense model (Llama3.3) across different batch sizes.}  \label{fig:sensitivity_batch}
    \vspace{-4mm}
\end{figure}

\cref{fig:sensitivity_batch} plots kernel-level speedup at a fixed 64/512
input/output context across batch sizes. The Qwen3-Next MoE slows by 2\% at
batch 16: at low batch, few experts activate, so few threadblocks launch and
the access ratio becomes imbalanced. The slowdown vanishes as the batch grows,
and \papername reaches roughly 6\% speedup at batch 256 and sustains it at 1024,
where all 512 experts activate (each decode token routes to 10 experts plus
one shared). The dense Llama-3.3 shows the opposite trend: speedup falls from
7\% at batch 8 to 6\% at batch 64, and turns into a 0.5\% slowdown at batch
256. Grace Hopper does not L2-cache loads from GPU to
host~\cite{fusco2024understanding}, so at large batch a weight tile is split
across threadblocks that compute subsets of the batch; inter-threadblock reuse
at L2 then matters, host loads bypass it, and \papername loses performance.
 
\noindent\textbf{Out-of-the-box (OOTB) models.}
We evaluate two more models to confirm these regimes: Qwen3-30B MoE and DeepseekCoder-33B~\cite{guo2024deepseek} dense model.
The MoEs incur slowdown at batch 16, where only a few experts
activate, and gain as the batch grows, reaching 3.7--6.2\% at batch 256.
The dense models gain at low batch sizes but start to incur slowdown at batch 256, where multiple
CTAs share a weight tile and host loads bypass the L2.
 
\subsection{Performance of DSR's Approximate CAP}
\label{sec:dsr_capp}
 
\begin{figure}[htb!]
    \centering
    \includegraphics[width=\linewidth]{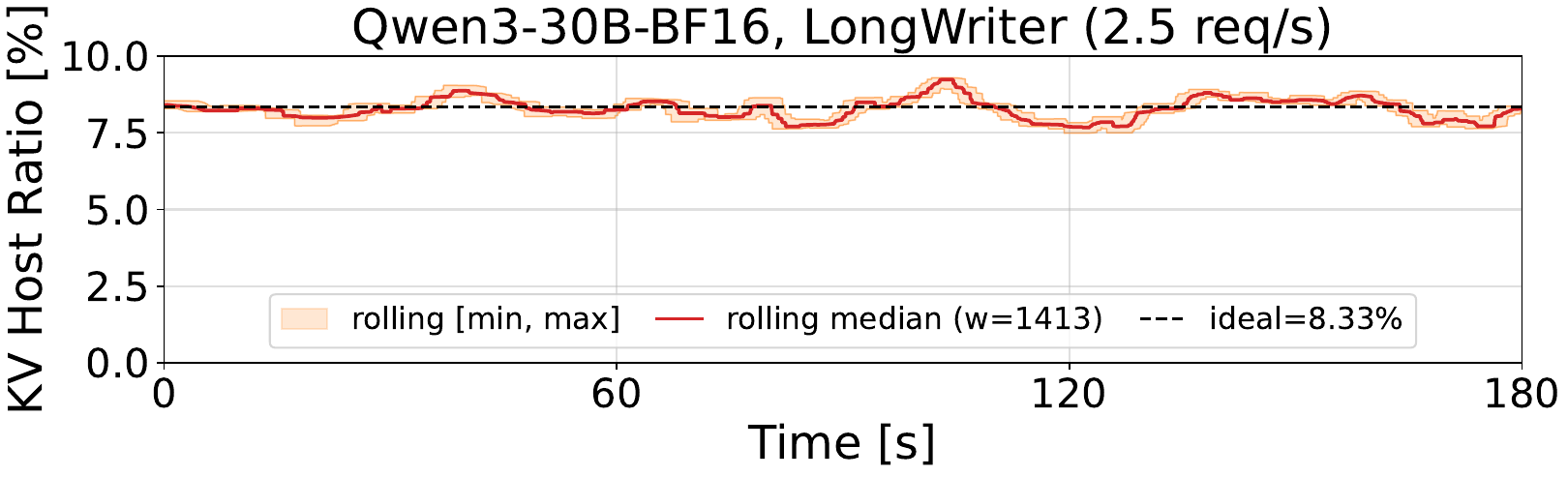}
    \vspace{-6mm}
    \caption{DSR keeps the host ratio of KV cache near ideal despite low number of attention heads (H=4).}    \label{fig:dsr_kv_ratio}
    \vspace{-4mm}
\end{figure}

DSR holds the host fraction of KV cache accesses near the ideal ratio at runtime, even when a model exposes few KV heads.
\cref{fig:dsr_kv_ratio} plots the realized host
fraction over a Qwen3-30B run (four KV heads) on LongWriter at 2.5 req/s,
against the configured ideal of 8.3\% (one of every 12 pages on the host tier, \cref{sec:methodology}). The rolling median stays between 7.7\% and 9.2\%
throughout, within $\pm1\%$ of the ideal ratio, and the rolling min--max band stays
equally narrow. The worst-case granularity bound is $1/H = 25\%$
(\cref{sec:design_impl}), yet the realized deviation is just 1\%, an order of magnitude below this bound, because the block-type mix of DSR spreads single-head-host blocks across
requests. DSR thus approaches the determinism of MPP for dynamic KV data,
despite the indirection of the free pool.
 
\subsection{Bandwidth Ratio Ablation with NVLinks}
\label{sec:bw_ratio_sens}
\begin{figure}[htb!]
    \centering
    \includegraphics[width=0.8\linewidth]{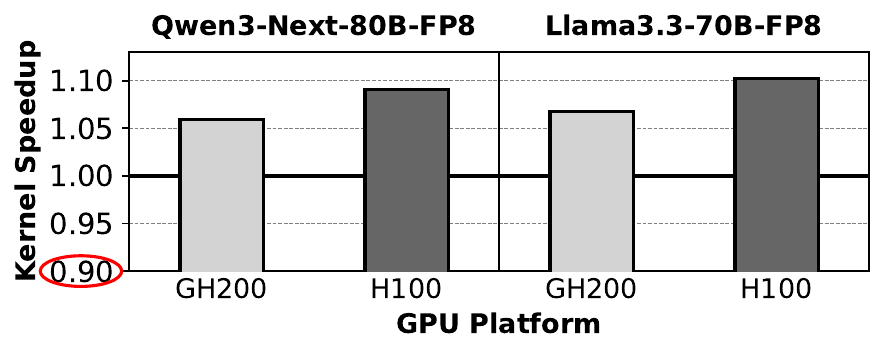}
    \vspace{-3mm}
    \caption{Speedup of expert kernel from a MoE model (Qwen3-Next) and matmul kernel from a dense model (Llama3.3) across different hardware platforms. Input activations use batch size of 256 (MoE) and 64 (dense). H100-emulation provides better speedup due to higher $\alpha$.}
    \vspace{-3mm}
    \label{fig:sensitivity_alpha}
\end{figure}

We emulate a higher bandwidth ratio ($\alpha=12\%$) with two NVLink-connected
H100 GPUs (80GB each); NVLink provides cache-coherent load-stores with latency
and caching behavior similar to the Grace CPU~\cite{fusco2024understanding}.
The LLM runs on one H100, and the second serves only its memory, measured at
2.81TiB/s HBM and 351GiB/s over NVLink. \cref{fig:sensitivity_alpha} compares
the expert (Qwen3-Next) and matmul (Llama-3.3) kernel speedups on GH200 and
the H100 emulation. \papername reaches 9.6\% average latency speedup on the H100
versus 6.3\% on GH200, because the H100 offers a higher $\alpha$. \papername thus
benefits systems with higher-bandwidth secondary tiers.

\subsection{Larger Models with Tensor Parallelism}
\begin{figure}[htb!]
    \centering
    \includegraphics[width=0.7\linewidth]{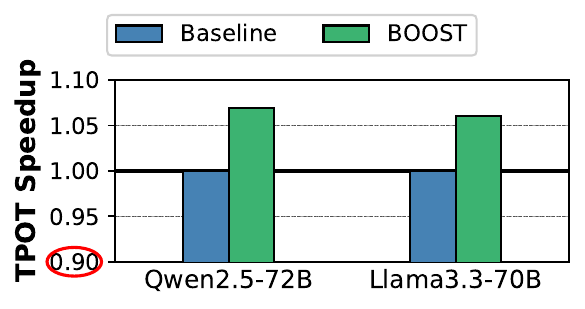}
    \vspace{-4mm}
    \caption{TPOT speedup normalized to HBM-only serving for larger models with tensor parallelism (TP=2) on H100-emulation. \papername improves TPOT by 6.5\% on average.}
    \vspace{-4mm}
    \label{fig:main_tp}
\end{figure}

Our GH200 node is single-GPU, so we extend the H100
emulation~(\cref{sec:bw_ratio_sens}) to evaluate scalability on larger models
with tensor parallelism (TP). \cref{fig:main_tp} shows TPOT speedup normalized
to HBM-only for Qwen2.5-72B (batch 6) and Llama3.3-70B (batch 4) at TP=2 for
low-latency serving, both in BF16. \papername improves TPOT by 6.5\% on average,
extending to multi-GPU systems.

\section{\revhl{Discussion}}

\subsection{Future Hardware Design}
\label{sec:future}
 
Scaling \papername to larger future batches needs two things. First, the GPU must
L2-cache host loads, as in prior works~\cite{beyond_socket, carve}, to retain
performance under high inter-threadblock reuse~(\cref{sec:batch_sens}). Second,
the SMs must saturate both tiers at once: on an H200, HBM saturation
uses 84\% of the SMs and the host tier needs another 20\%, exceeding
the SM budget, whereas an H100 needs only 62\% for HBM. Future
systems must provision enough SMs and buffers to drive both tiers.

\parhead{Performance Model.} To project \papername onto future systems, we
build an analytical throughput model. Rather than slow cycle-level
simulation or coarse roofline bounds, it models LLM operators at
tens-to-hundreds of microseconds, shards their FLOPs and bytes by
parallelization scheme, and maps them onto hardware. It captures
heterogeneous memory through a critical-path timing model with
\textit{slack} analysis that accounts for prefetching across L2 and HBM. We
implement it in Python.
We validate it against published GB200
NVL72 measurements of DeepSeek-V3/R1 decode~\cite{sglang_gb200} (2K context,
HBM-only, DP+EP). It predicts peak per-GPU decode throughput
within $-$6.0\% in FP8 (8,539 vs 9,087 tok/s/GPU) and $+$4.9\% in a mixed
FP4 configuration with FP8 attention (14,041 vs 13,386 tok/s/GPU).

\begin{figure}[!htb]
    \centering
 \includegraphics[width=0.75\linewidth]{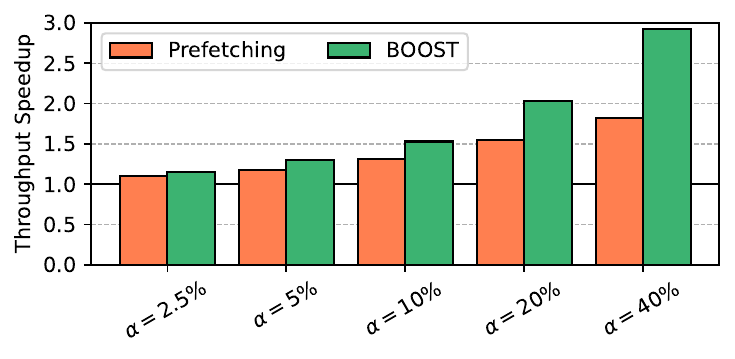}   
    \vspace{-3mm}
    \caption{Projected normalized throughput speedup of \papername and prefetching for GB200 running DeepSeek R1 FP4 at 128K context/user at varying $\alpha$ (high-TPS setting). \papername outperforms prefetching by up to 60\%.}
    \label{fig:sensitivity_study_c2c_bandwidth}
    \vspace{-5mm}
\end{figure}

The Grace Blackwell (GB200)~\cite{gb200_whitepaper} pairs two GPUs per CPU
($\alpha=3\%$); future Vera-class systems~\cite{vera_rubin} reach 900GB/s C2G
bandwidth ($\alpha$ up to 44\%). \cref{fig:sensitivity_study_c2c_bandwidth}
projects \papername and idealized prefetching on GB200 for DeepSeek
R1~\cite{guo2025deepseek} in FP4 at 128K context (TP=2), scaling $\alpha$ from
2.5\% to 40\%, normalized to HBM-only. \papername gains rise with $\alpha$,
approaching 3x and exceeding prefetching by up to 60\%, so \papername scales to
future systems.

\subsection{Applicability of \papername}
\label{sec:applicability}

\parhead{Generality.}
\papername remains useful even when the working set fits entirely in HBM, unlike prefetching which lowers demand bandwidth.
The host tier holds idle bandwidth regardless of capacity pressure, and \papername converts this bandwidth into a 4.3\% iso-batch TPOT gain~(\cref{fig:main_combined}); the added capacity then arrives as a further benefit when workloads grow~(\cref{sec:cap_bw_isolation}).
\papername also requires no kernel changes, so unmodified models benefit out of the box~(\cref{sec:batch_sens}).
Finally, our design assumes only a cache-coherent secondary tier, making BOOST extensible to other tiered memory systems.

\parhead{Decoding Regimes.} \papername targets the memory-bandwidth-bound decode instances, and two
regimes limit its gains. First, \papername degrades when many CTAs share one weight
tile, because host loads bypass the L2~\cite{fusco2024understanding} and break the reuse.
Hopper caps the MLP output tile at $N{=}128$, so beyond batch 128 (256 on
Blackwell) multiple CTAs share a tile, as in prefill. \papername therefore confines
placement to decode, placing a fraction of weights on the host tier under
co-location; L2 caching of host loads~\cite{beyond_socket, carve} would lift
this limit~(\cref{sec:future}). Second, \papername loses at small MoE batches, where
few experts activate and too few CTAs launch to hide the latency of
host-routed CTAs: Qwen3-Next slows 2\% at batch 16~(\cref{fig:sensitivity_batch}), but
the slowdown vanishes with scale, and \papername gains 6\%
at the kernel level at batch 1K. To
sustain consistent gains, we recommend decode batch sizes of
64--1K for MoE and
8--128 for dense models (up to 256 on Blackwell owing to dedicated TMEM for output tiles).

\parhead{Interconnects.} \papername is functional wherever the GPU issues
load-stores to a host tier. Our H100 emulation~(\cref{fig:sensitivity_alpha}) realizes a
multi-GPU NVLink configuration, where one GPU serves memory for another. \papername
also runs on PCIe, but its low bandwidth ratio ($\alpha \approx 2\%$) yields
marginal gains, as we expect for CXL. High-bandwidth NVLink-attached and
disaggregated memory are the strongest fit on future high-$\alpha$
systems~(\cref{sec:future}).

\subsection{\papername Design Considerations}
\label{sec:design_considerations}
 
\parhead{System integration.} \papername places its mechanisms in the serving
system because that layer alone sees the access pattern of each kernel and the
allocation of each tensor that CAP requires; the OS is agnostic to kernel
boundaries (\cref{sec:os_cuda}), and cuBLAS~\cite{cublas} admits no kernel
edits. \papername modifies only the weight loader and KV-pool manager, assigns
pages through \texttt{mmap} and \texttt{mbind}, and performs no migration:
PAM places all pages statically while the kernels issue demand loads, and DSR
changes only the assignment of KV data to those pages at
runtime~(\cref{sec:design_impl}). It adds no GPU fragmentation, since vLLM
partitions HBM into fixed KV blocks at startup~\cite{vllm}, and reserves 9.6GB
(2\% of CPU capacity) in 1GB pages.

\parhead{Requirements for transparent CAP.} \papername obviates kernel
changes by relying on three properties of LLM kernels. First, cache-coherent
load-stores let the SM reach the host tier with ordinary loads, so \papername needs
no message passing or in-kernel orchestration. Second, high-performance
kernels such as paged attention and tiled GEMMs address virtual tensors and
ignore physical placement, so \papername reassigns pages through \texttt{mmap} and
\texttt{mbind} and reconstructs the same tensor without changing the access
pattern. Third, tiling, per-CTA resource use, and TMA produce a steady per-CTA
access stream, so \papername places each CTA on one tier and routes an $\alpha$
fraction to the host tier. \papername does not split a CTA across tiers, which
stalls the SM~(\cref{sec:insights}); it interleaves across CTAs at page
granularity.
The handwritten kernel of \cref{sec:mpp} beats \papername by 0.2
points (7.3\% versus 7.1\%), too little to justify editing various kernels
and models against the single 800-LoC change of \papername.

\section{Related Work}
\label{sec:related}

\subsection{Page Management for GPU Memory System}
Agarwal et al.~\cite{agarwal2015unlocking} migrate CPU data to HBM with
one-touch access and nearby-page prefetching, regulating aggressiveness by
bandwidth utilization. Such migration reduces to linear placement, which slows
our add kernel by 72\% (LPP). BATMAN~\cite{chou2017batman} controls migration
direction to hold a target local-access rate, preserving proportionality. LLM
inference has predictable access patterns, and \textit{any} migration spends
HBM bandwidth on writes, limiting gains; \papername migrates nothing and places
pages directly. Agarwal et al.~\cite{agarwal2015page} show full bandwidth needs
bandwidth-proportional placement (BAPP-R). We are the first to show that
both proportionality \textit{and} concurrency are required, which BAPP-R fails
to provide.

\subsection{CPU Tiered Memory Management}
Nimble~\cite{yan2019nimble} improves page-migration throughput in the OS
kernel. Adaptive Migration Policy~\cite{heo2020adaptive} selects a migration
policy from application behavior. Soar~\cite{liu2025tiered} guides migration
with loaded latency and memory-level parallelism beyond page hotness.
Colloid~\cite{vuppalapati2024tiered} places pages proactively to minimize
loaded access latency, and HeteroOS~\cite{kannan2017heteroos} extends
cooperative placement into the hypervisor to cut migrations. These policies
suit latency-sensitive CPUs with unknown access patterns, but they treat memory as a hierarchy rather than peers, so they incur migration
overhead and do not provide bandwidth-proportional placement, similar to serving systems.

\subsection{GPU Memory Capacity Expansion}
FlexGen~\cite{sheng2023flexgen} and ZeRO-Offload~\cite{ren2021zero} offload
model weights and optimizer state to CPU memory for training on
memory-constrained GPUs. For inference, MoE-Lightning~\cite{cao2025moe-lighning},
PRESERVE~\cite{yuzuguler2025preserve}, Pie~\cite{xu2024pie}, and
others~\cite{kong2025serving, yu2025taming, xue2025moeinfinity,
chen2025ktransformers} prefetch weights and KV data to HBM, scheduling
migration to overlap with computation; others offload to
SSD~\cite{pan2025instattention} or speculate~\cite{zhuge2025specoffload}, and
MoonCake~\cite{qin2025mooncake} pools CPU memory across nodes for warm KV state
such as prefix cache. These schemes expand capacity but treat the tiers
hierarchically and expend HBM write
bandwidth. \revhl{They target the compute-bound high-throughput regime,
where spare compute hides migration latency.} \papername instead treats the
tiers as peers with SM-issued cache-coherent loads, \revhl{achieving additive bandwidth in the bandwidth-bound online
regime, where no compute slack exists.} \revhl{InfiniGen~\cite{lee2024infinigen}
and NEO~\cite{jiang2025neo} are orthogonal, reducing the volume of KV fetched
and offloading attention computation to the CPU respectively, while \papername
accelerates the demand traffic across both tiers.}

\subsection{LLM Serving Systems}
Frameworks such as vLLM~\cite{vllm}, TRT-LLM~\cite{trtllm}, and
SGLang~\cite{zheng2024sglang} layer many optimizations. Kernel
optimizations---FlashAttention~\cite{dao2022flashattention,
dao2023flashattention, shah2024flashattention},
FlashDecode~\cite{hong2024flashdecoding++}, FlashInfer~\cite{ye2025flashinfer},
operator fusion~\cite{zhuang2024mononn, nayak2024fusemax, chimera,
zhai2023bytetransformer}, and megakernels~\cite{wu2024mirage}---tile caches and
cut memory traffic and launch overhead, raising compute efficiency; they
complement \papername, which raises bandwidth utilization. Scheduling optimizations
such as continuous batching~\cite{yu2022orca}, prefill
chunking~\cite{agrawal2024taming}, and KV migration~\cite{sun2024llumnix}, and
memory schemes such as LLM-aware allocation~\cite{zhang2025jenga} and
PagedAttention~\cite{pagedattention_vllm}, remain HBM-centric.
vAttention~\cite{prabhu2025vattention} removes non-contiguous virtual memory
via CUDA VMM and complements \papername: it removes the reshaping in DSR while PAM
makes page management CAP-compliant.

\section{Conclusion}
LLM decode is memory-bandwidth-bound, yet current serving systems underutilize the host memory bandwidth. Extracting additive bandwidth of both tiers requires
\textit{concurrent and bandwidth-proportional accesses (CAP)} within every
wave, which existing placement and migration schemes do not provide. We present
\papername, a runtime system that performs CAP through wave-aware page placement
for static weights and runtime KV-pool management for dynamic KV data, without
kernel changes. \papername improves TPOT by 4.3\% where prefetching
degrades it by 6\%, and raises throughput by 31\% on average---with 4\% bandwidth-improvement that only BOOST's concurrent access provides---15\% above prefetching.

\begin{acks}
We thank the reviewers of ISCA-2026 and MICRO-2026 for their valuable feedback, and Prof. Tushar Krishna for providing access to Grace Hopper node through Georgia Tech's CRNCH Cluster for initial experiments. Anish was partly funded by Nvidia Graduate Fellowship. This work was supported by NSF grant 233304.

\end{acks}

\bibliographystyle{ACM-Reference-Format}
\bibliography{refs}

\end{document}